%% file: concept.tex
\documentclass[12pt, letter]{emulateapj}
\usepackage{graphicx,amssymb,amsmath,epstopdf, xfrac,subfigure}
\pdfoutput=1

\newcommand{\asec}{{^{\prime\prime}}}
\newcommand{\Reff}{{$R_{\rm e}$}}

\newcommand{\msun}{M$_{\odot}$}
\newcommand{\Msun}{{\rm M}_{\odot}}
\newcommand{\sersic}{S\'{e}rsic}

\newcommand{\N}[1]{N$_{#1}$}
\newcommand{\per}{$^{-1}$}

\newcommand{\atl}{ATLAS$^{\rm 3D}$}

\newcommand{\specialcell}[2][c]{%
  \begin{tabular}[#1]{@{}c@{}}#2\end{tabular}}

\newcommand{\lspecialcell}[2][c]{%
  \begin{tabular}[#1]{@{}l@{}}#2\end{tabular}}

\newcommand{\NII}{[\ion{N}{2}]}

\newcommand{\OIII}{[\ion{O}{3}]}
\newcommand{\SII}{[\ion{S}{2}]}

\shorttitle{{MaNGA}: Mapping Nearby Galaxies at APO}
\shortauthors{Bundy et al.}

\submitted{}

\begin{document}


\title{Overview of the SDSS-IV  M{\rm {\footnotesize a}}NGA Survey: \\ Mapping Nearby Galaxies at Apache Point Observatory}

\slugcomment{version 5.3 (accepted for publication in ApJ)}

\input{authors.tex}



\begin{abstract}

  We present an overview of a new integral field spectroscopic survey called MaNGA (Mapping Nearby Galaxies at Apache Point
  Observatory), one of three core programs in the fourth-generation Sloan Digital Sky Survey (SDSS-IV) that began on 2014 July 1.
  MaNGA will investigate the internal kinematic structure and composition of gas and stars in an unprecedented sample of 10,000
  nearby galaxies.  We summarize essential characteristics of the instrument and survey design in the context of MaNGA's key
  science goals and present prototype observations to demonstrate MaNGA's scientific potential.  MaNGA employs dithered
  observations with 17 fiber-bundle integral field units that vary in diameter from 12\arcsec\ (19 fibers) to 32\arcsec\ (127
  fibers).  Two dual-channel spectrographs provide simultaneous wavelength coverage over 3600--10300 \AA\ at $R$$\sim$2000.  With a
  typical integration time of 3 hr, MaNGA reaches a target $r$-band signal-to-noise ratio of 4--8 (\AA\per\ per 2\arcsec\
  fiber) at 23 AB mag arcsec$^{-2}$, which is typical for the outskirts of MaNGA galaxies.  Targets are selected with $M_* \gtrsim
  10^9~\Msun$ using SDSS-I redshifts and $i$-band luminosity to achieve uniform radial coverage in terms of the effective radius,
  an approximately flat distribution in stellar mass, and a sample spanning a wide range of environments.  Analysis of our
  prototype observations demonstrates MaNGA's ability to probe gas ionization, shed light on recent star formation and quenching,
  enable dynamical modeling, decompose constituent components, and map the composition of stellar populations.  MaNGA's
  spatially resolved spectra will enable an unprecedented study of the astrophysics of nearby galaxies in the coming 6 yr.

\vspace{0.2cm}

\end{abstract}


\section{Introduction}\label{sec:intro}

The study of galaxy formation is ultimately about understanding the processes that formed a heterogeneous universe from a
homogeneous beginning.  This field encompasses an enormously rich range of different physical phenomena, from the growth of primordial
fluctuations in the early universe, through the cooling and condensation of gas into molecular clouds and stars, to the formation
of supermassive black holes associated with the emission of copious radiation as they grow within galactic bulges. Major advances
in our understanding of galaxy formation owe a great deal to highly successful galaxy surveys conducted with the Sloan Telescope
at low redshift \citep{york00, strauss02} and with a variety of facilities at higher redshift.  These have not only begun to map the cosmic
distribution of galaxies, but also revealed how the star formation rates (SFRs), metallicities, stellar populations, morphologies, and
black hole growth rates of galaxies vary with mass, environment, and time.

Despite their success, all spectroscopic surveys larger than a few hundred galaxies suffer from a significant drawback: they
sample a small  subregion of each targeted galaxy (e.g., defined by the location of the light-collecting fiber or the orientation of the slit) to derive
{\em integrated} spectroscopic properties, or, at best, one-dimensional (1D) gradients along single position angles.  The complex and rich internal structure of galaxies is largely missed, and in
the worst case, inferred measurements may be biased by the fact that they sample only the galaxy's center or a particular axis.

By ``peering inside'' galaxies with integral field spectroscopy, a whole new dimension becomes available for addressing key
questions about the formation and evolution of galaxies, and the richness of this dimension only grows as the wavelength coverage
increases.  The projected spatial distribution of strong emission lines and key diagnostics of stellar populations can reveal the
sites of new and recent star formation and distinguish proposed quenching mechanisms.  We can use signatures of stellar ages,
element abundances, and star formation histories combined with dynamical decomposition to understand how disks grew and evolved
and how pressure-supported components (bulges and ellipticals) assembled over time.  Full, two-dimensional (2D) dynamics allows us to
chart the distribution of angular momentum and construct mass models that reveal how forming baryonic components interacted and
influenced their host dark matter halos.  Additionally, the dynamically inferred distributions of baryons can be independently
verified with spectrophotometric indicators of stellar mass, based on surface-gravity estimates calibrated for the age,
metallicity, and abundance of the stellar populations---possible only with broad spectral coverage.

This promise of spatially resolved spectroscopy has motivated many observations with integral field units (IFUs).  Over the past
decade, the pioneering work of the SAURON \citep[72 E/S0/Sa galaxies,][]{dezeeuw2002} and ATLAS$^{\rm 3D}$ surveys \citep[260 E/S0
$K$-band selected galaxies\footnote{{\tt http://purl.org/atlas3d}}, volume-limited within 42 Mpc or $z\la
0.01$,][]{cappellari2011} heralded a new era of IFU surveys at $z \approx 0$ and introduced a novel way to classify early-type
galaxies based on their IFU kinematics rather than morphology. They also demonstrated the power of detailed dynamical modeling of a
large sample of galaxies with IFU kinematics.  More recently the DiskMass Survey \citep{bershady10} has targeted 146 face-on,
star-forming disk galaxies using fiber-IFU observations at high spectral resolution ($R \sim 10,000$).  As a counterpart to the
dynamical studies of early-type galaxies with the SAURON instrument, the DiskMass Survey provided insight on the internal mass profiles
and mass-to-light (M/L) ratios of disk populations in galaxies including those much like the Milky Way, finding strong evidence,
for example, that most disk galaxies are ``submaximal'' with dark matter fractions greater than $\sim$50\% out to 2.2 disk scale
lengths \citep{bershady11}.  

A number of other recent campaigns targeting local galaxies complement the surveys described above.  The VENGA survey
\citep[VIRUS-P Exploration of Nearby Galaxies,][]{blanc13a} prioritized high spatial resolution, deep observations of 30 nearby
spirals chosen for their extensive ancillary multiwavelength data.  VENGA is particularly well suited to studying star formation
and the interstellar medium \citep[ISM; e.g.,][]{blanc13}.  The SLUGGS Survey \citep[SAGES Legacy Unifying Globulars and
GalaxieS,][]{brodie14} utilizes Subaru/Suprime-Cam imaging combined with Keck spectroscopy to obtain integral field data for 25 nearby
early-type galaxies that reach $\sim$8 \Reff.  Taking advantage of globular cluster tracers, SLUGGS combines multiple probes of
the kinematics and chemical composition of galaxies at very large radii.  Meanwhile, the MASSIVE Survey \citep{ma14} focuses on a
sample of $\sim$100 of the most massive galaxies within 108 Mpc, addressing the assembly, stellar populations, and black hole
scaling relations of galaxies in a mass regime that has been poorly studied so far ($M_* \gtrsim 10^{11.5}$ \msun).

New IFU facilities on large telescopes and in space are providing new opportunities to obtain resolved spectroscopy at higher
redshifts.  These include KMOS \citep[K-band Multi-Object Spectrograph,][]{sharples06} and the recently commissioned MUSE
instrument \citep[Multi Unit Spectroscopic Explorer,][]{bacon10} as well as IFU concepts for 30~m class telescopes, the James Webb
Space Telescope \citep[MIRI: Mid-Infrared Instrument][]{wright04}, and WFIRST \citep{spergel13}.  The rise of such facilities
indicates that the future of high-$z$ galaxy studies will be dominated by integral field observations, and yet, despite the
significant efforts described above, low-$z$ samples remain heterogeneous and too small to provide a complete, local benchmark to
which high-$z$ results may be compared.

Even today, the context for interpreting high-$z$ studies remains unclear. For example, $z \sim 2$ observations have emphasized
the surprising diversity of kinematic structure, including large rotating disks, compact dispersion-dominated objects, and major
mergers \citep[e.g.,][]{shapiro08, forster-schreiber09, law09}, not to mention evidence for winds and outflows
\citep[e.g.,][]{steidel10,genzel11}.  High velocity dispersions dominate the kinematics, even for the largest galaxies with the
most disk-like velocity fields. Many of the galaxies have highly irregular morphologies and host giant kpc-sized ``clumps'' of
young stars \citep{forster-schreiber11,law12}.  It is unclear whether these phenomena represent fundamentally different modes of
assembly and star formation compared to present-day galaxies \citep[see][]{green10}.

Several recent efforts are now under way (\S\ref{sec:comparison}) to construct comparison low-$z$ IFU benchmarks that can answer
such questions as well as exploit the power of resolved spectroscopy obtained for large statistical samples to address major
unsolved questions in galaxy formation regarding the assembly, processing, recycling, and dynamical redistribution of baryons
inside dark matter halos. An initial important step was taken by the ongoing CALIFA\footnote{\tt http://califa.caha.es} survey
\citep[Calar Alto Large Integral Field Area,][]{sanchez12}, which aims to observe a sample of 600 galaxies with $\sim$1 kpc
resolution and a wavelength range from 3750 to 7000 \AA\ using the PPak IFU developed for the DiskMass Survey. CALIFA celebrated
its first public data release of 100 galaxies in October 2013 and has observed over 400 galaxies to date. The more recently
launched SAMI\footnote{\tt http://sami-survey.org} survey \citep[utilizing the Sydney Australian Astronomical Observatory
Multi-object Integral Field Spectrograph,][]{Croom2012} is making major gains using a multiplexed fiber-IFU instrument with two
wavelength channels (3700--5700 \AA\ and 6250--7350 \AA) and relatively high spectral resolution at $\sim$7000 \AA\ to target 3400
galaxies at a spatial resolution of 1--2 kpc. The SAMI instrumentation and sample selection are described in \citet{bryant14}
and a description of its Early Data Release is given in \citet{allen15}. 

Mapping Nearby Galaxies at Apache Point Observatory (MaNGA) represents another advance, surveying 10,000 galaxies across a wide
dynamic range in $M_*$, environment, and SFR with uniform radial coverage. Especially important is MaNGA's long,
continuous wavelength coverage, which enables full model fitting over the many spectral features present between 3600 and 10300
\AA.  MaNGA's unique wavelength coverage in the near-IR is of prime importance in this respect thanks to diagnostics that are
sensitive to the initial mass function (IMF), such as \ion{Na}{1} and the \ion{Ca}{2} triplet \citep{conroy12}, as well as tracers of cooler photospheres
that can constrain thermally pulsating asymptotic giant branch (TP-AGB) stars and other intermediate-age populations
\citep{maraston05}. While ionized gas emission line diagnostics are more discrete, there are many valuable lines from
[\ion{O}{2}]$\lambda$3727 to [\ion{S}{3}]$\lambda$9531 that in concert will help break degeneracies and improve diagnostic
calibrations on gas-phase abundance indicators.

In this paper we provide an overview of the MaNGA concept, from instrumentation and hardware to sample design and early
proof-of-concept observations with prototype instrumentation that were obtained through a generous donation of observing time by
the SDSS-III Collaboration \citep{eisenstein11}.  The MaNGA survey launch date was 2014 July 1, and the program will run for 6
yr, utilizing half of the dark time available in the fourth-generation Sloan Digital Sky Survey (SDSS-IV).

We begin with the scientific motivation for MaNGA and its key questions in \S\ref{sec:science}.  We then present a
``back-of-the-envelope'' design for an IFU survey that achieves these science goals and delivers a sample of many thousand
galaxies (\S\ref{sec:envelope}).  We demonstrate how MaNGA is ideally suited to perform such a survey.  A brief overview of the MaNGA
instrumentation, including a description of the prototype hardware, is given in \S\ref{sec:instrument}.  Full details on
instrument design, testing, and assembly will be given by N.~Drory et al.~(in preparation).  D.~A.~Wake et al.~(in preparation) present
the sample design, optimization, and final selection of the survey---this is summarized in \S\ref{sec:survey}.  We discuss how the
MaNGA design compares quantitatively with other IFU surveys in \S\ref{sec:comparison}.  Finally, \S\ref{sec:results} presents the
prototype MaNGA observations already obtained (``P-MaNGA'') and highlights several early results that demonstrate MaNGA's
scientific promise.

Additional forthcoming papers will establish the complete technical description of the survey, including a description of the
science requirements, integrated survey systems and operations, and a characterization of the initial data quality; MaNGA's
observing strategy, including dithering and image reconstruction in the face of variable conditions and differential atmospheric
refraction (DAR); the survey's software framework, metadata tracking and repositories, and reduction pipeline; and MaNGA's approach to
flux calibration.

Throughout this work, we use the AB magnitude system \citep{oke83} and adopt a standard cosmology with $H_0$=100 $h$ km s$^{-1}$
Mpc$^{-1}$, $\Omega_M$=0.3 and $\Omega_{\Lambda}$=0.7.  For estimates of stellar mass ($M_*$) we adopt the IMF from \citet{chabrier03}.

\section{Science Goals}\label{sec:science}

The $\Lambda$CDM framework provides a cosmological context for galaxy formation that posits that dark matter halos grow from the
``bottom up,'' hierarchically assembling into increasingly massive structures with time.  The fundamental components of
present-day galaxies formed as a result of various complex processes that act on the baryons residing in these evolving halos.
Observations over the past two decades have identified $z \sim 1$--3 as the epoch during which this growth peaks.  At subsequent
times, star formation and assembly gradually decline in a global sense, while subpopulations experience possibly different
evolutionary histories depending on their mass, gas content, and the halos in which they live.  Mergers and star formation, fueled
by the accretion of fresh gas and likely regulated by feedback from various mechanisms, continue to drive growth as the
morphological mix of populations evolves with time and environment.  Eventually, star formation shuts down in what is observed to
be a mass-dependent fashion.

MaNGA seeks the {\em physical origin} of the mechanisms that drive this evolution.  What are the processes that shape the assembly
of bound components, regulate growth via star formation, merging, and quenching, and affect the chemical abundance of both stars
and gas?  MaNGA will shed light on the nature of gravitational collapse, gas infall and dissipation, star formation and its
regulation through feedback mechanisms, galaxy merging, secular evolution, accretion onto supermassive black holes, and how all of
these phenomena connect with the environment in which galaxies live. MaNGA's strength in addressing these questions arises from the
statistical power of a large sample of galaxies probing a wide range of environment, where each galaxy is resolved into 2D maps of
stellar and gas composition and kinematics. MaNGA's unparalleled sample size, combined with its ability to probe the internal
structure of galaxies from the near-UV to the near-IR, enables transformative discoveries and surprises.

\bigskip

\noindent{\bf MaNGA's Key Science Questions:}
\begin {enumerate} 

\item How are galaxy disks growing at the present day and what is the source of the gas supplying this growth?  {\em (Section
    \ref{sec:growth})}

\item What are the relative roles of stellar accretion, major mergers,
  and secular evolution processes in contributing to the present-day growth of
  galactic bulges and ellipticals? {\em (Section \ref{sec:growth})}

\item How is the shutdown of star formation regulated by internal processes within galaxies and externally-driven processes that
  may depend on environment? {\em (Section \ref{sec:quenching})}
  
\item How is mass and angular momentum distributed among different components and how has their assembly affected the components
  through time?
  {\em (Section \ref{sec:components})}

\end {enumerate}

\subsection{Growth and Assembly}
\label{sec:growth}

Significant uncertainty remains in understanding how galaxies accrete gas within their large-scale environment, particularly in
terms of the relative importance of ``hot'' versus ``cold'' accretion modes. MaNGA will map disk fueling and recent gas accretion
by measuring spatially dependent SFRs and gas-phase metallicities that can be directly compared to stellar
metallicities of the older population. Gas accreting along ``cold'' filaments will tend to have high angular momentum and
close-to-primordial metallicities \citep[e.g.][]{spitoni11, mott13}. Such accretion manifests itself through abrupt drops in
metallicity at large radii \citep{moran12}.  These features have been detected in small samples of galaxies at high redshift
\citep{cresci10,troncoso14}.  In contrast, accretion from surrounding hot coronae produced by supernovae and active galactic nucleus (AGN) ejecta yield star
formation with a more even spatial distribution in the disk, often concentrated toward the center, and with relatively high
metallicities \citep[e.g.][]{bresolin12,spitoni13}.  Recently, \citet{sanchez14} argued for a universal gas-phase metallicity
gradient, with 10\% of their sample harboring central ``holes'' and the majority exhibiting a flattening in the outskirts
($\sim$2 \Reff) that may be driven by secular processes.  Distinguishing and understanding the contributions from these different
processes to the evolving galaxy population as a whole requires a statistical approach that utilizes MaNGA's large sample, which
will establish how metallicity gradients depend on galaxy mass {\it as a function of galaxy environment}.

In addition to gas accretion followed by star formation, major mergers are another contributor to present-day galaxy growth.  The
degree and nature of merging activity remain highly uncertain, however.  MaNGA will provide a full 2D {\em kinematic} census of
merging galaxies in a mass-limited sample, opening a unique view on minor interactions and merger phases that are difficult to
detect in imaging alone.  Through comparisons with suites of hydrodynamical merger simulations \citep[e.g.,][]{lotz10, lotz10a},
new statistical metrics can be devised that will help diagnose the merger phase (i.e., first approach, second passage, final
coalescence) and the mass ratio of the interacting galaxies.

\begin{figure*}[htb!]
\centering
\includegraphics[width=0.95\textwidth, angle=0]{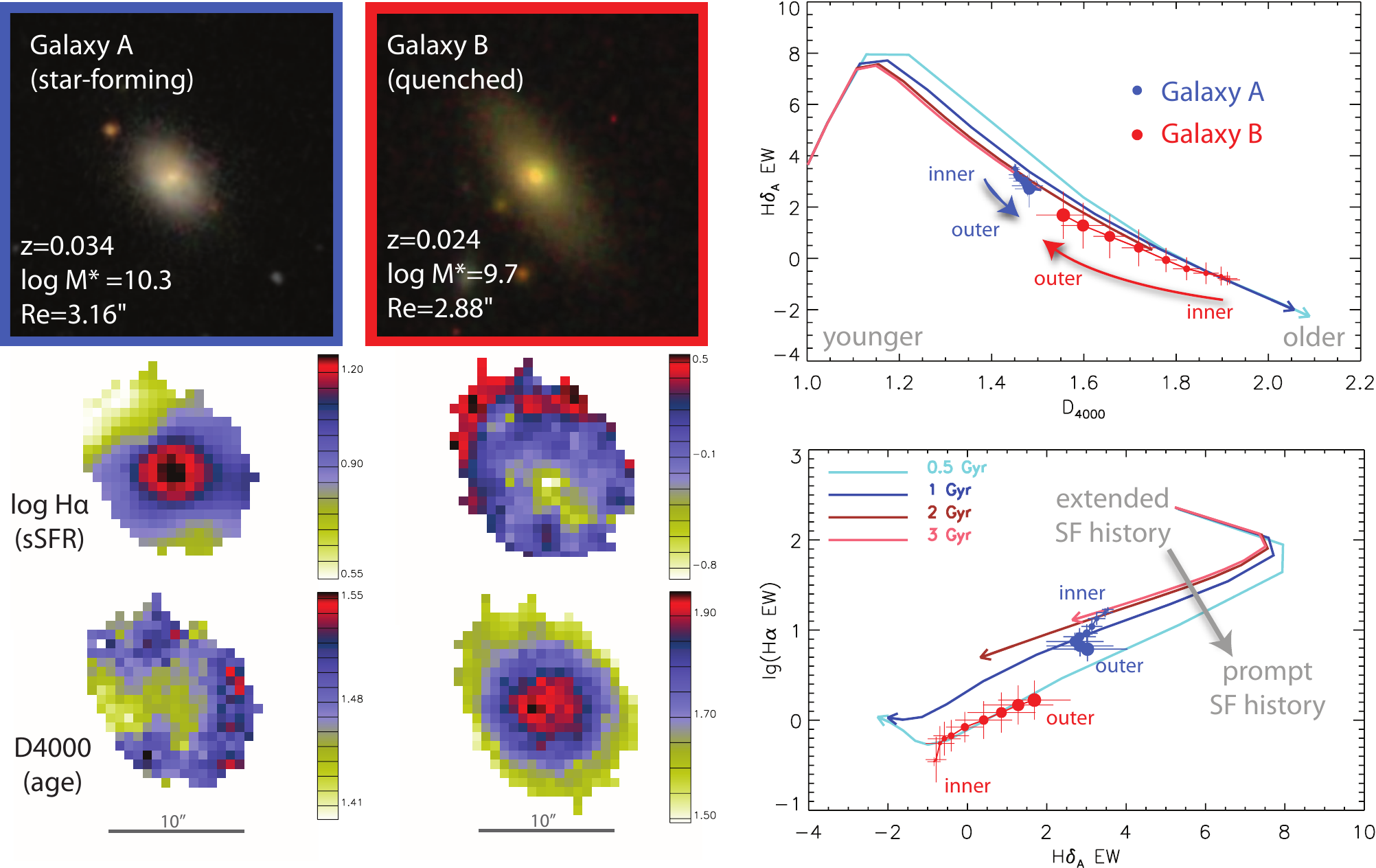}
\caption{Illustration of MaNGA's ability to reveal radial variations in the recent star formation history of target galaxies.  The figure shows preliminary results from C.~Li
  et al.~(in preparation) for two galaxies observed during the 2013 January prototype run (see \S\ref{sec:obs}). The two left
  columns present a star-forming (p9-19E) and quenched (p9-19D) galaxy.  From top to bottom, we show the SDSS images, maps of the
  rest-frame H$\alpha$ EW (a proxy for the specific star formation rate, sSFR), and maps of the D4000 index (a proxy for stellar
  age). The rightmost column presents two panels comparing the rest-frame H$\delta_A$ EW, H$\alpha$ EW, and D4000 diagnostics against one
  another as measured in different radial bins in each galaxy. The curves represent models with SFRs $\propto \exp(t/\tau)$, each
  with a different value of $\tau$ as labeled. For
  the star-forming galaxy, the inner region has a more extended and recently active SFH.  The quenched galaxy, on the other hand,
harbors a central region dominated by a seemingly older stellar population compared to that residing in the outer regions of this
galaxy.  All EW measurements are in \AA.}

\label{fig:quenching}\label{fig:cheng}
\end{figure*}

Mergers are often invoked to explain the evolving morphological mix of galaxy populations and, more recently, the size growth
observed in elliptical galaxies \citep{van-dokkum08}.  Direct measures, however, suggest that the merger frequency may be too low
\citep{bundy09, newman12}, motivating further tests of merger-driven scenarios.  If minor mergers increase sizes by building
extended wings \citep[e.g.,][]{naab09}, for example, outer components should be older and more metal poor \citep{lackner12}.  In
the small samples observed to date, stellar age gradients are negative but shallow among disk galaxies
\citep{sanchez-blazquez14} and flat among early-types \citep[e.g.,][]{mehlert03,
  Sanchez-Blazquez_2006, kuntschner10}, although trends consistent with late-time growth are seen at larger radii, near 2 \Reff\
\citep{greene13}.  Even the bulges in galaxy centers may be composite, with a fraction of stars formed in the disk and driven in
by secular processes, in situ star formation from preprocessed disk gas \citep{johnston14}, and a third component of
stars accreted in mergers \citep[e.g.,][]{drory07}.  More generally, it appears that various processes conspire to produce
``inside-out'' star formation histories, with earlier, more rapid formation occurring in the centers of galaxies, at least with masses above $M^*$
\citep[e.g.,][]{perez13}.  Via comparisons to models that can predict the fraction of stars formed through various channels as a function of
galaxy mass \citep[e.g.,][]{zavala12}, MaNGA will provide definitive answers to the question of how much material is ``accreted''
versus formed ``in situ'' in galaxies with varying structure, mass, and environment.

\subsection{Quenching and Environment}
\label{sec:quenching}

The array of physical mechanisms responsible for quenching star formation in galaxies is one of the most hotly debated questions in galaxy
formation.  Standard explanations invoke galaxy mergers that can drive inflows of cold gas that fuel central starbursts and
AGNs.  With a large fraction of the gas consumed by the starburst, outflows powered either by winds
generated by the accreting black hole or by the starburst itself \citep[e.g.,][]{sharp10} can heat the remaining gas and perhaps even expel it from the
galaxy altogether \citep[see][]{hopkins08}.  Alternatively, the heating may occur {\em outside} the galaxy.  This could take the
form of tidal harassment from a nearby companion or, if the galaxy is a satellite in a group or cluster, could be related to the
hot gaseous medium surrounding it.  In these environments, internal gas supplies may be stripped away \citep[e.g., ram-pressure
stripping, see][]{moran07} or prevented from replenishing (e.g., ``starvation'', \citealt{mccarthy08, bekki09}).  Meanwhile,
previously quenched systems in low-density environments may experience rejuvenation \citep[e.g.,][]{thomas10}.  

Signatures of recent star formation quenching can be correlated with other galaxy properties to understand the relevant physical
mechanisms at work.  MaNGA has unique statistical power in this regard because its effective \'etendue in the near-UV (see
\S\ref{sec:hardware}) allows precise measurements of spectral indices like D4000 and H$\delta$. These
indicators gauge the relative prominence of $\sim$1 Gyr old populations compared to the fraction of younger stars ($t <
100$ Myr) indicative of recent bursts of star formation \citep[e.g.,][]{kauffmann03a}.  Additional spectral features can further
constrain stellar ages and the makeup of the stellar population.  The TiO features redward of 6000 \AA\ are a signature of AGB
stars \citep{maraston11}, and the CN bandheads redward of 7000 \AA\ indicate the presence of carbon stars \citep{maraston05}, both populations
associated with ages of $\sim$1 Gyr helpful in breaking age/metallicity degeneracies. 

Gradients in these diagnostics of recent star formation history will be used to build a comprehensive picture of the quenching
process. A demonstration of MaNGA's potential in this regard is given by a preliminary analysis of prototype MaNGA observations
taken from Li et al.~(in preparation), with highlights shown in Figure \ref{fig:cheng}. Quenching from an AGN or nuclear starburst
exhibits an inside-out behavior (e.g., the globally quenched ``Galaxy B''), and AGNs may be further identified by the ionization
state of gas in the central regions (\S\ref{sec:ion}) and the presence of winds. The importance of mergers and interactions in
triggering quenching can be statistically deduced by correlating observed ``quenching gradients'' with the smoothness of stellar
and gas velocity fields. A process like ``morphological quenching,'' in which secular bulge growth alters the global potential in
a way that stabilizes the gaseous disk against fragmentation \citep{martig09}, imprints flatter gradients and dynamical
signatures, for example.

MaNGA's large sample will enable thorough investigations of quenching processes that are related to environment.  Studies of
satellite quenching in group-scale halos suggest processes that occur over 1--3 Gyr timescales \citep[e.g.,][]{weinmann09,
  lackner12a}, but MaNGA will also capture a valuable sample of rare, but more dramatic, phenomena, including tidal stripping in
action.  These observations will reveal (1) whether stripping-induced truncation events are ubiquitous in dense cluster regions, (2) the
typical radii at which the truncation is occurring, and (3) the dependence of truncation radius on galaxy properties, halo mass, and
cluster-centric radius.

\subsection{Formation of Galaxy Subcomponents}
\label{sec:components}

\begin{figure*}[htb!]
\centering
\includegraphics[width=0.95\textwidth, angle=0]{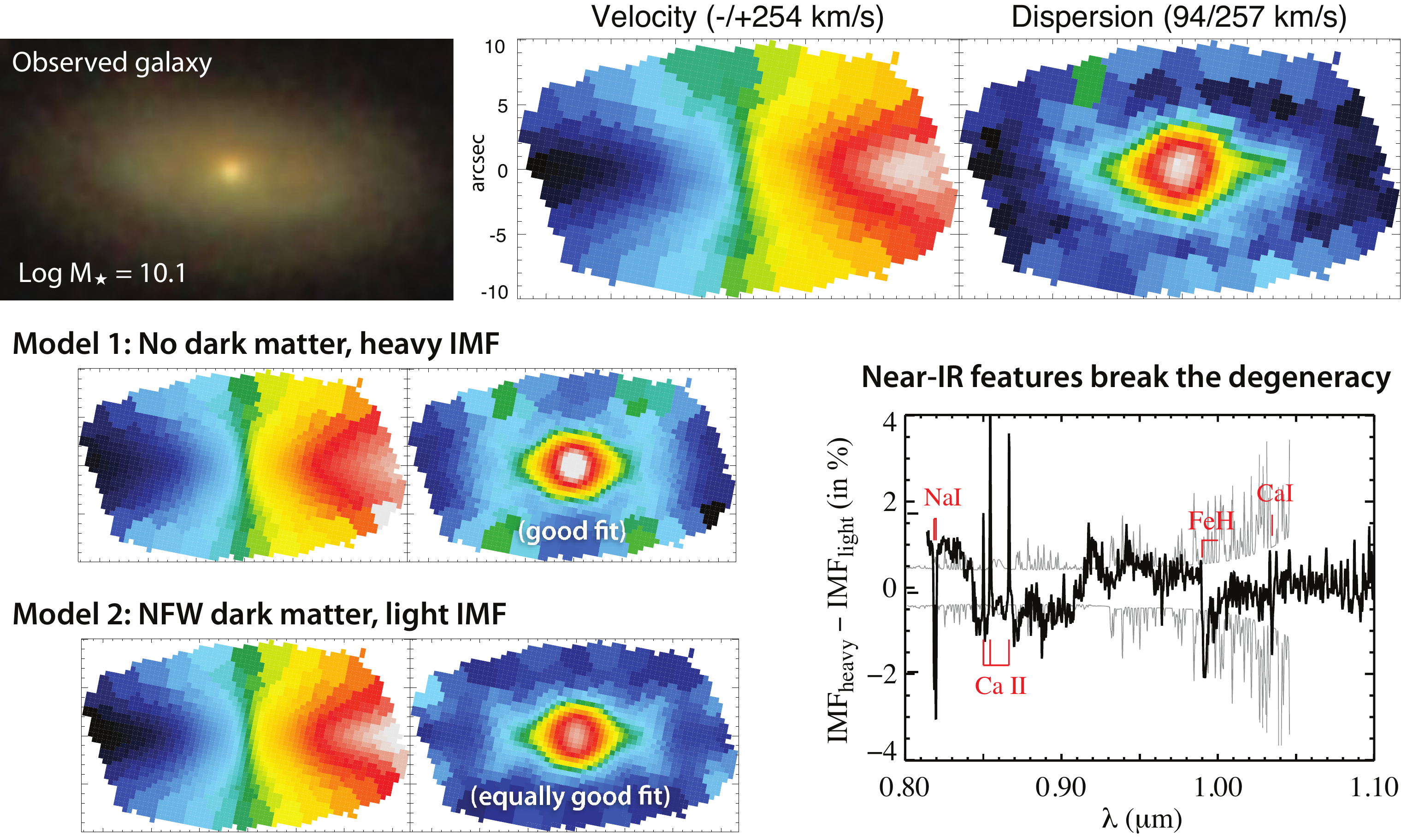}
\caption{MaNGA dynamical and stellar population constraints on galaxies. The top row shows the observed SDSS $gri$ image, velocity
  field, and velocity dispersion for a galaxy in the 2013 January MaNGA prototype run (p11-61A). The bottom left panel presents
  two equally good model fits; Model 1 assumes that the mass is dominated by stars, with a ``bottom-heavy'' IMF,  while Model 2
  assumes a Navarro-Frenk-White (NFW) dark matter halo \citep{navarro97} with a ``bottom-light'' IMF. Absent spectroscopic IMF diagnostics, the two models are
  degenerate. MaNGA will be able to break the degeneracy by measuring several IMF-sensitive indicators, as shown in the bottom right panel. The
  black line displays the differences between the two IMF models assumed here, while the gray envelope shows the estimated MaNGA
  errors for a spectral stack of the inner regions of a galaxy.  The \ion{Na}{1} and \ion{Ca}{2} features are firmly detected.
  Marginal detections of variation in FeH and \ion{Ca}{1} could be improved with further stacking.}
\label{fig:imf}\label{fig:remco}
\end{figure*}

The coevolution of the mass and angular momentum distributions of the disk, bulge, and dark matter halo components of galaxies
ultimately define their present-day sizes and dynamical states. Different formation channels from dissipative collapse to fueling
by cold streams, as well as tidal forces from mergers or instabilities, have varying effects on these distributions. If they can
be decomposed from one another we can learn about the galaxy's formation history. This ``fossil record'' remains poorly explored,
however, in large part because uniform measures of 2D kinematic fields have not been possible for large samples. Even when such
observations are available, dynamical measures constrain the {\em total} mass distribution, posing a fundamental challenge to
decomposing galaxies into their constituent parts without making various assumptions, including the shape of the dark matter
distribution and the stellar mass-to-light ratio.  Yet, it is these very quantities that hold the clues to the galaxy's
formation history.

The unknown contribution of dark matter to the total mass budget within the optical extent of galaxies is degenerate with the
factor of $\sim$2 uncertainty in the stellar mass-to-light ratio caused by uncertainties in the IMF. Resolving this degeneracy
would shed light on the largely unknown influence assembling baryons have on the dark matter halo, while measurements of the
stellar M/L ratio and shape of the IMF would provide valuable new insight on the conditions in which stellar disks and spheroids
formed. Since these conditions vary with mass, gas properties, redshift, and environment, these measurements constrain the physics
governing star formation and assembly over cosmic time, and would reveal potential IMF variations, which are assumed to be
constant by nearly every current study of galaxy evolution.

MaNGA has the unique potential to break the degeneracy between mass and light by combining state-of-the-art dynamical modeling
\citep[e.g,][]{cappellari06, cappellari2008, van-den-bosch08, de-lorenzi09, long10} {\it and} IMF-sensitive spectrophotometric
modeling \citep[e.g.,][]{conroy2012, la-barbera13, spiniello14} for the first time. This capability is enabled by MaNGA's
two-dimensional mapping of stellar rotation and velocity dispersion as well as its long wavelength coverage, which enables
simultaneous measurements of widely distributed, intrinsically narrow (and therefore weak) absorption features.  The near-IR
wavelength range in MaNGA survey data will provide unique access to gravity-sensitive features \citep{conroy2012} that constrain
the faint-end of the IMF ($<$0.5M$_\odot$), and hence the stellar mass-to-light ratio ($\Upsilon$*), while at the same time the
extended spectral coverage provides strong constraints on the star formation history, metallicity, and abundance of the stellar
population.  For an illustration of MaNGA's ability to address this issue, see Figure \ref{fig:remco}.  Both the spectral and
dynamical approaches currently employ rather restrictive assumptions in order to measure the stellar $M/L$, dark matter and
IMF. The spectral approach has currently been restricted to a few population components, with homogeneous chemical composition and
IMF. The dynamical approach generally assumes simple geometries for the galaxy, as well as the shape and profile for the halo. The
combination of both approaches, employing radically different assumptions, has the potential to break the degeneracies in the
models to uniquely constrain the stellar $M/L$ and dark matter content. This synergy can provide fundamental insights in our
understanding of dark matter and stellar populations in galaxies.

Turning to late-type galaxies with younger stellar ages and dynamically ``colder'' components, addressing similar questions
requires a different approach.  In addition to obtaining definitive measures of scaling relations like Tully--Fisher
\citep[e.g.,][]{masters06, reyes11}, MaNGA's unique strength in providing multiple kinematic tracers across a long,
uniformly calibrated wavelength range raises a potentially transformative but more speculative prospect.  Because these different
gas and stellar tracers are sensitive to different populations as a function of their age and composition, the stellar velocity
dispersion from the disk's self-gravity may be inferred by measuring the asymmetric drift (AD) between stellar and gas tangential
speeds. Not only does this information provide a unique dynamical estimate of disk mass, but by isolating older and younger stellar
populations, the dynamical heating rate of disks can be determined. The expected AD signal is $\sim$10-20 km s$^{-1}$ based on the Milky
Way and observations from the DiskMass Survey \citep{bershady10}, well within MaNGA's reach at even modest signal-to-noise ratio. The
capacity for innovation in dynamical analysis with MaNGA data is greatly enhanced by the uniformity of the observations and data
analysis. Success will not only provide a windfall in our understanding of local disk galaxies but set the stage for
application of this method in surveys at higher redshift.

Finally, with a statistical approach, it will be valuable to compare mass estimates from multiple methods over a variety of length
scales.  MaNGA will provide dynamical masses, $M_{\rm dyn}$, within $\sim$$1.5 R_e$ as described above and more robust estimates
of $M_*$.  Next-generation HI surveys, such as Apertif \citep{verheijen08} shallow and/or medium-depth surveys at Westerbork, and
WALLABY (the ASKAP HI All-Sky Survey, B.~S.~Koribalski et al., in preparation) at lower
declination, can provide rotation curves out to much larger radii. Remarkably, at even greater distances, overlap with deep
panoramic imaging (e.g., using the Hyper Suprime-Cam on the Subaru Telescope) can provide weak-lensing mass estimates of the dark
matter halo $M_{\rm halo}$ with factor of $\sim$2 uncertainties for stacks of several hundred MaNGA galaxies.  Comparing these
different mass estimators opens a regime for testing modifications to general relativity designed to explain cosmic acceleration
\citep[e.g.,][]{zhao10} that predict differences in the lensing mass, $M_{\rm lens}$, compared to $M_{\rm dyn}$, or potentially
even in the mass traced by stars compared to gas as a result of different screening processes \citep[see][]{jain13}. Deviations would be strongest for
isolated, low-mass galaxies ($M_* = 10^9$--$10^{10}$ \msun) where $M_{\rm dyn}$ may be as much as 30\% larger than $M_{\rm lens}$.

\subsection{Sample Size}
\label{sec:discovery}

\begin{figure*}[ht!]
\epsscale{1.0}
\plotone{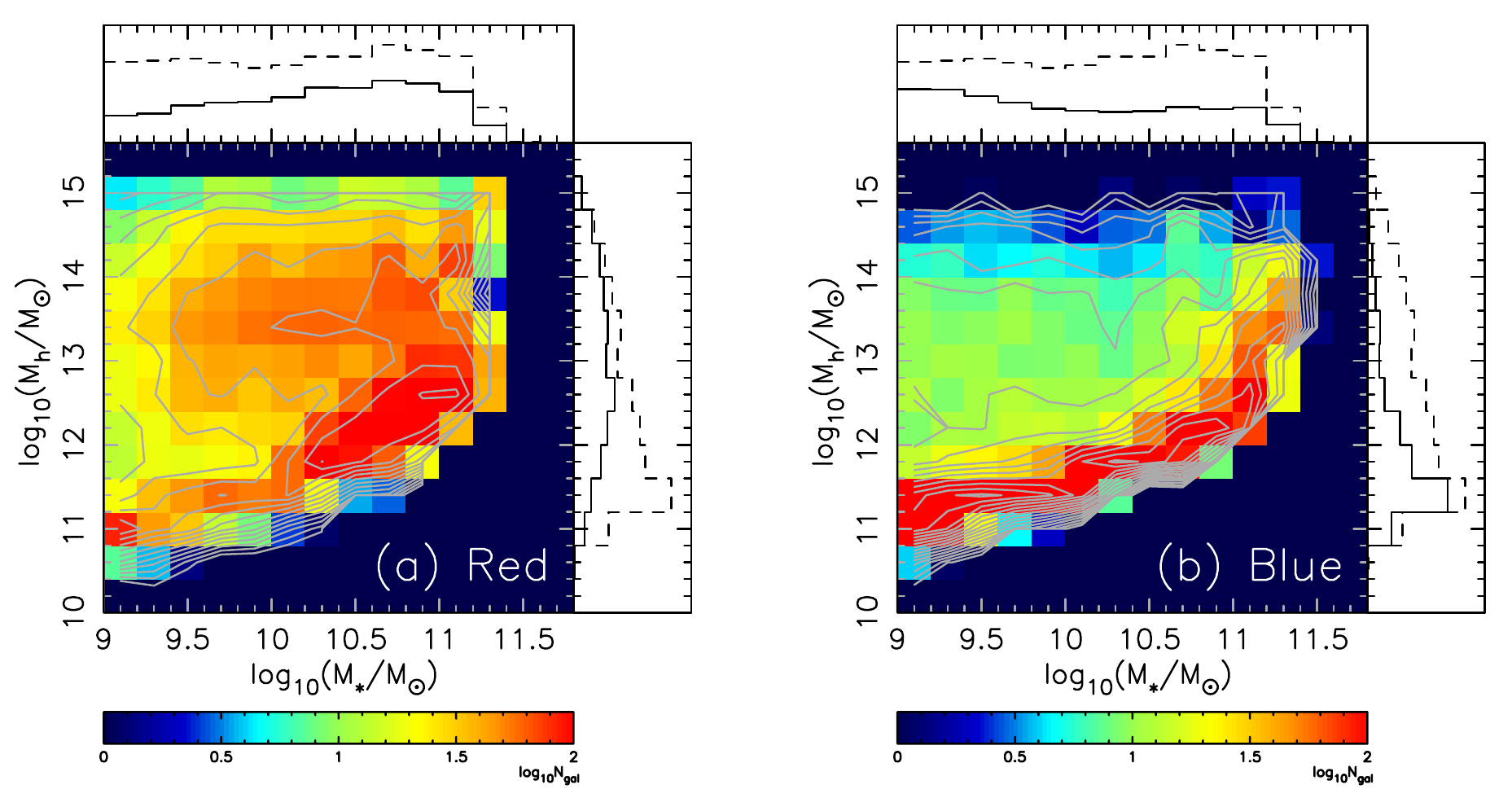}
\caption{Expected number of MaNGA galaxies as a function of both $M_*$ ($x$-axis) and $M_{\rm halo}$ ($y$-axis) for both red
  galaxies (left) and blue galaxies (right) based on mock catalogs informed by the semianalytic model of \citet{guo11} and
  constructed as in \citet{li06}. Ten to $\sim$100 galaxies are commonly reached in bins (each spanning 0.2 dex in $M_*$ and 0.4 dex
  in $M_{\rm halo}$) across a large dynamic range in both quantities, demonstrating
  the statistical power of MaNGA's 10k sample.  Normalized histograms show 1D marginalized $M_*$ distributions (top axes) and
  $M_{\rm halo}$ distributions (right axes), with dashed lines indicating the full primary sample and solid lines corresponding to
  red (left) or blue (right) galaxies.
  \label{fig:mocks}}
\end{figure*}

We finally discuss additional motivation for MaNGA's sample size, which, when compared in terms of uniform spatial coverage
and resolution, is roughly an order of magnitude larger than any current or planned survey (\S\ref{sec:comparison}). Obviously, a
larger sample is always desired, but a simple binning argument suggests that 10,000 is a valuable target. This value arises
because a general investigation would demand approximately six bins across each of the three ``principal components'' that define
the galaxy population: stellar mass, SFR (or morphology), and environment (e.g., halo mass). The number of
galaxies required in each bin can be estimated by assuming the common situation in which the single-measurement precision is
roughly equal to the expected difference in signal from bin to bin. For example, MaNGA aims for a precision in derived stellar age
gradients of $0.1$~dex per decade in \Reff, which is on par with the typical variation in this quantity among galaxies studied to
date \citep{mehlert03,kuntschner10, spolaor10}. In such cases, the significance of a detected difference between bins is equal to
$\sqrt{n/2}$, where $n$ is the number of galaxies in each bin. If we demand a 5$\sigma$ detection, we arrive at a requirement of
$n=50$. Multiplying this number by the 6$^3$ bins required to sample the galaxy parameter space returns a sample size of 10,800. A
more detailed mock-up of MaNGA's ability to explore the galaxy population parameter space is shown in Figure \ref{fig:mocks}.
These arguments demonstrate MaNGA's statistical power to explore new trends in the galaxy population.

The large sample is also needed to study rare populations such as mergers, AGNs, post-starbursts,
galaxies with strong outflows, etc., especially new classes of phenomena that can only be detected with resolved
spectroscopy. MaNGA will provide statistically meaningful samples (i.e., more than 100 objects) of subpopulations that occur with
a frequency of just a few percent and will detect handfuls of even rarer galaxy types.

\section{Back-of-the-Envelope Design}\label{sec:envelope}

Before describing the MaNGA instrumentation and survey design in detail, we present a ``back-of-the-envelope'' concept
for a MaNGA-like survey that demonstrates how our key design components depend on a handful of assumptions and
constraints.  We will assume the use of fiber-bundle IFUs with a fiber diameter matched to
typical ground-based seeing, $\sim$2\arcsec.  A reasonable S/N requirement is 5--10 \AA$^{-1}$ per fiber in the outskirts of the
target sample, which we adopt to be 1--2 \Reff.  Since the on-sky, apparent surface brightness of galaxies is roughly conserved to
within 50\% to $z \sim 0.1$, a single exposure time will achieve the desired S/N for a large fraction of ``nearby'' galaxies. For
the BOSS spectrographs \citep{smee13} on the Sloan 2.5m telescope \citep{gunn06}, the required exposure time is about 3 hr.

Maximizing the physical spatial resolution demands target galaxies with low redshifts.  We can quantify a limiting spatial resolution by considering the
typical intrinsic sizes of galaxies which vary roughly from 3 to 9 kpc.  To make an IFU observation worthwhile, one needs to sample the
smallest galaxies with at least three radial bins, which sets a spatial resolution requirement of $\sim$1 kpc.  Kinematic features
and interesting spatial structures (e.g., bars, bulges, spiral arms) also have scales that are $\sim$1 kpc in size.  Assuming typical ground-based
seeing (and $z < 0.2$) the physical scale, $\delta$, subtended by a 2\arcsec\ spatial element is $\delta \approx 35z$~kpc.
Setting $\delta = 1 $kpc yields a target redshift of $z \approx 0.03$.

Further design constraints originate from the desired sample size of $\sim$10,000 galaxies. The volumetric number density of galaxies
with $M > 10^{9}$~\msun in the local universe is about 10$^{-2}~{\rm Mpc}^{-3}$. Assuming an available targeting area of 3500
deg$^2$, just less than half the size of the SDSS-I \citep{york00}, a redshift limit of $z > 0.03$ is required to have {\em enough} volume
from which to draw a 10k sample. From these constraints we obtain an on-sky targeting density with $z > 0.03$ of 3 deg$^{-2}$,
which corresponds to $\sim$20 IFUs for the Sloan 3$^{\circ}$ diameter field of view. With the 3 hr exposure time and $\approx$500
pointings (plates), completing the survey requires 190 nights. Factoring in the weather (50\% clear), dark time (40\%), and time sharing
with eBOSS (50\%), we arrive at a 6 yr survey duration.

Finally, we address the fiber budget and distribution in IFU sizes. At $z \sim 0.03$ the typical optical radius of a galaxy (1.5
\Reff) is about 10\arcsec. This radius would be subtended by five 2\arcsec\ fibers, which, if hexagonally packed, would sum to a
61-fiber IFU. With this average IFU size and 20 IFUs needed to sample the SDSS field of view, the total number of fibers required
is about 1300.  This is slightly greater than the 1000 fibers in the BOSS instrument and shows why it was desirable to increase
MaNGA's fiber budget to 1423 (\S\ref{sec:blocks}).  Turning to the IFU size, in order to study gradients and maps across the
galaxy population as a function of \Reff, uniform coverage is required to a target radius based on \Reff. Since \Reff\ varies by
an order of magnitude across the galaxy population, the ideal complement of IFUs should also vary by an order of magnitude in
size. This prospect, however, yields a range in the number of fibers per IFU of two orders of magnitude (since the number scales
as the area subtended by the IFU). Thus, the majority of the available fiber budget, a number fixed by the size of the
spectrograph, would be assigned to just a few large bundles.  To mitigate this problem, it is clear that the target selection
should make use of the decreasing apparent angular diameter of galaxies with increasing redshift, at the sacrifice of some
physical spatial resolution (in kpc).  The intrinsically largest galaxies should be targeted at somewhat greater distances
compared to intrinsically smaller galaxies.  With a redshift range of $0.03 < z < 0.10$ the required dynamic range in IFU size is
compressed to roughly a factor of three and an order of magnitude in terms of number of fibers per IFU.

This basic design captures the motivation for the more detailed specifications of the MaNGA Survey as
summarized below.  We are fortunate that the hardware constraints from IFU technology and the SDSS spectrograph and
telescope design specifications are well matched to the requirements of an idealized ground-based IFU survey of a
large sample on a 2.5 m telescope.  The large field of view is a key advantage.  In addition, we have 
optimized the remaining design elements by coupling the instrumentation to the sample selection and observing strategy.

\begin{figure*}[ht!]
\centering
\includegraphics[width=0.80\textwidth, angle=0]{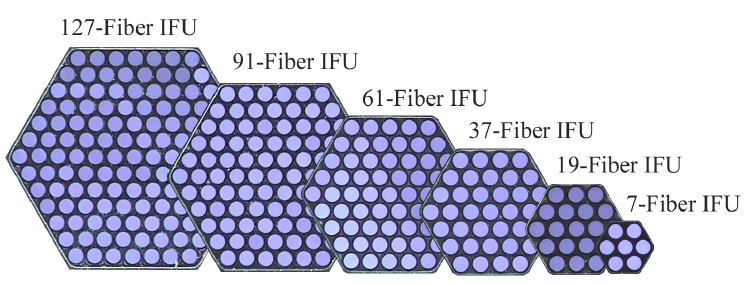}
\caption{Actual images of six MaNGA IFUs from the first batch of survey-ready components.  The photographs have been cropped to
  each IFU's hexagonal area and overlaid in this presentation.  An example of each science IFU is shown, as well as a seven-fiber
  ``mini-bundle'' (far right), which is used for flux calibration.  The hexagonal packing is extremely regular---the positions of
  individual fibers (120 $\mu$m core diameter, 2\arcsec\ on the sky, with an outer diameter of 150 $\mu$m) within the IFU deviate from their ideal locations by less
  than 3 $\mu$m.  The MaNGA IFU complement (Table \ref{tab:hardware}) is $12 \times $\N{7}, $2 \times $\N{91}, $4 \times $\N{37}, $4
  \times $\N{61}, $2 \times $\N{19}, and $5 \times $\N{127} per cartridge.}
\label{fig:ifu}
\end{figure*}

\section{Instrumentation and Facilities}\label{sec:hardware}\label{sec:instrument}

The MaNGA instrument (Drory et al., in preparation) utilizes the 2.5m Sloan Telescope in its spectroscopic mode as described in
\citet{gunn06}.  In its final form, MaNGA will provide 17 fiber-bundle science IFUs that can be deployed to target sources anywhere
within the 3$^{\circ}$ diameter focal plane. These IFUs feed light into the two dual-channel BOSS spectrographs \citep{smee13}, which
maintain the same configuration as was used in SDSS-III.  We provide summary information on both the prototype instrument and the
final specifications of the instrumentation used for ``production.''  Please see Drory et al.\ for details.

\begin{figure}
\centering
\includegraphics[scale=0.45, angle=0]{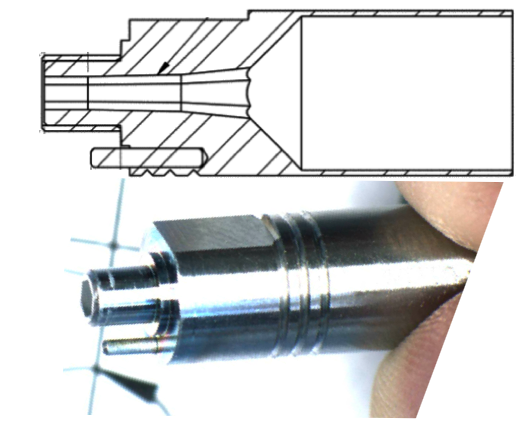}
\caption{MaNGA IFU ferrule concept.  The inner diameter tapering and transition to a hexagonal form are indicated by the arrow on
  the cutaway design drawing (top).  A clocking pin determines the positional angle orientation when the ferrule is plugged into
  the plate. The outer diameter, roughly 0.5 cm, makes the ferrules easy to handle and plug by hand.}
\label{fig:ferrule}
\end{figure}

\begin{figure}[]
\centering
\includegraphics[width=0.55\textwidth, angle=0]{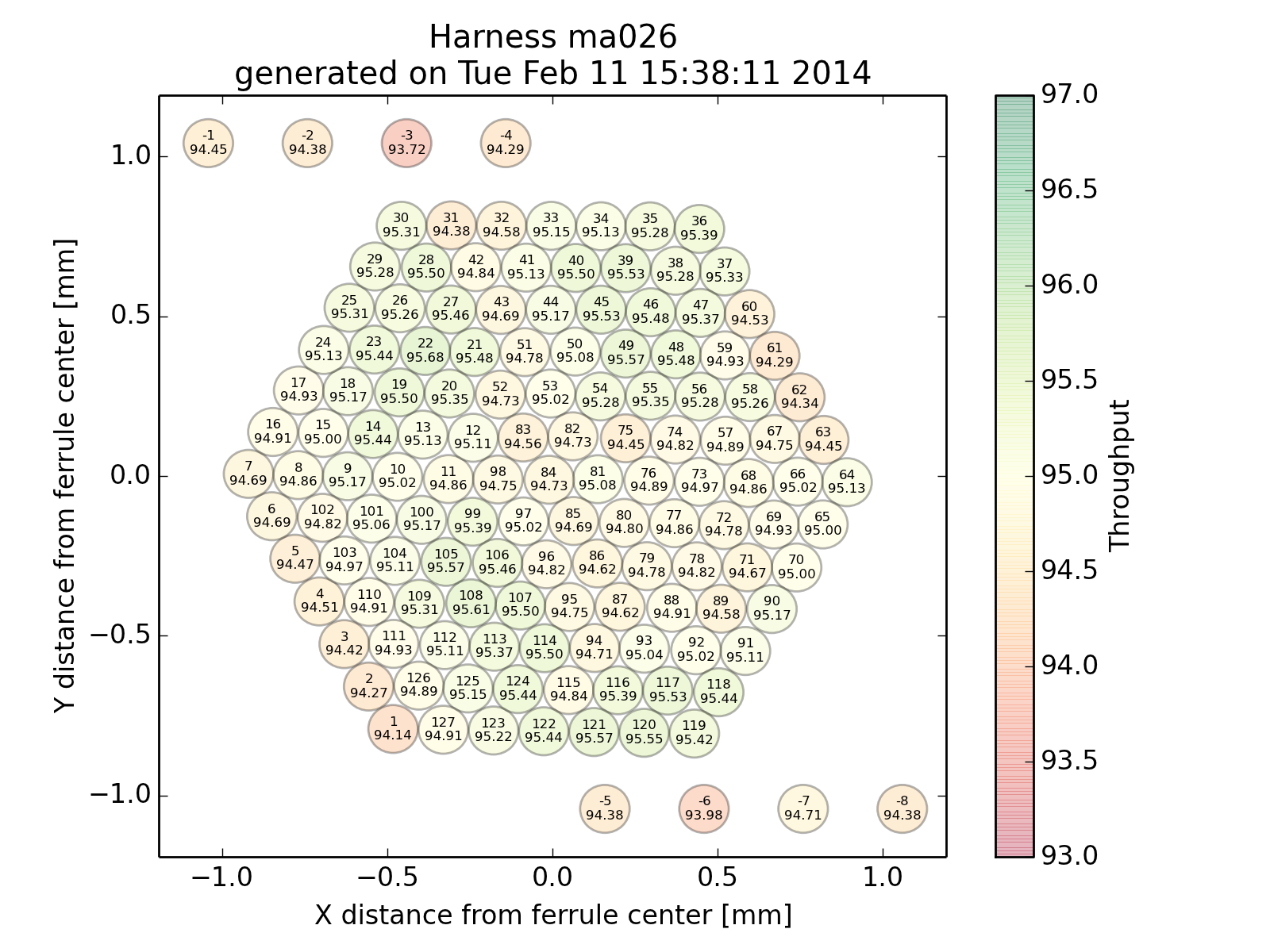}
\caption{Lab-measured ``throughput map'' for a survey-ready \N{127} IFU, generated automatically as part of our quality assurance
  testing and tracking.  This IFU is typical, with throughput values of $95$\%$ \pm 2$\%.}
\label{fig:throughput}
\end{figure}

\subsection{Fiber-bundle Integral Field Units}

The MaNGA IFU design took inspiration from the 61-core ``hexabundles'' developed and tested at the University of Sydney
\citep{bland-hawthorn10, bryant11, bryant14} and implemented in the SAMI instrument \citep{Croom2012}.  Several important differences in the
development path MaNGA pursued are worth noting, however.  First, the optical fiber used in hexabundles is first stripped of its outer
buffer and then ``lightly fused'' using an etching and heating process that effectively removes some of the fiber cladding and
thereby increases the packing density. The goal is to increase the fill factor---the ratio of live-core collecting area to total
area subtended by the bundle---but only to the point where light losses from the induced focal-ratio degradation (FRD) become
important. SAMI hexabundles reach a fill factor of $\sim$75\%.

For the MaNGA instrument, we desired as simple and inexpensive production process as possible, given that more than 200 separate IFUs
(distributed over multiple cartridges---see below) were required.  We therefore chose to maintain the protective fiber buffer,
which makes handling and assembly much easier and less costly, and also enables us to manufacture multiple IFU sizes using the same procedures.
MaNGA science IFUs range from 19 to 127 fibers in size, with diameters between 12$\asec$ and 32$\asec$ on-sky.  Examples of the
first production-level IFUs are shown in Figure \ref{fig:ifu}.  

As detailed in N.~Drory et al.~(in preparation), MaNGA's major technological advance is the design of the metal ferrule housing that holds the fibers
in place (Figure \ref{fig:ferrule}).  The inner hole begins as a wide circle that allows the bundle of fibers to be easily inserted
by hand.  The circular opening quickly tapers, and as it does, the circular shape gently becomes hexagonal.  The now hexagonal
opening narrows further until it reaches the final inner diameter, set to be slightly larger than what is needed for perfect fiber
packing in a hexagonal array.  Thanks to ``electrical discharge machining'' (EDM) techniques, fabrication tolerances of $\sim$3
$\mu$m can be achieved so that when fibers are inserted (by hand), even large bundles self-organize into a hexagonal array that is
remarkably regular.  Measured at the IFU end, the typical fiber is shifted by less than $\sim$3 $\mu$m rms (less than 2\% compared
to the outer diameter) from its position in a perfect hexagonal pattern.  Easy to implement and reproduce, this process produces
IFUs with $\sim$56\% fill factor.  Simulations of dithered observations demonstrate that the regular packing provides significant
gains in the uniformity of both the exposure depth and recovered spatial resolution across each IFU.

The 1423 fibers in each MaNGA cartridge have 2$\asec$ (120~$\mu$m) cores and are composed of high-performance, broadband fused silica with thin
cladding and buffer (120/132/150).  An antireflective coating applied to all fiber surfaces after IFU assembly increases the
average transmission by $\sim$5\% and makes the fiber-to-fiber throughput more uniform (N.~Drory et al., in preparation).  We have
successfully built early prototypes of MaNGA IFUs at the University of Wisconsin, but the large number of IFUs required working
with a commercial vendor. All the results described here are based on IFUs that were assembled, glued, and polished by
CTechnologies of New Jersey. Lab tests indicate excellent performance in the production-level IFUs, with no evidence for
design-related FRD beyond that of the SDSS single-fiber feed and typical throughput of 95\% $\pm$ 2\% (see Figure
\ref{fig:throughput}).

\subsection{Fiber Output Pseudo-slit and V-groove Blocks}\label{sec:blocks}

\begin{figure}
\centering
\includegraphics[scale=0.25, angle=0]{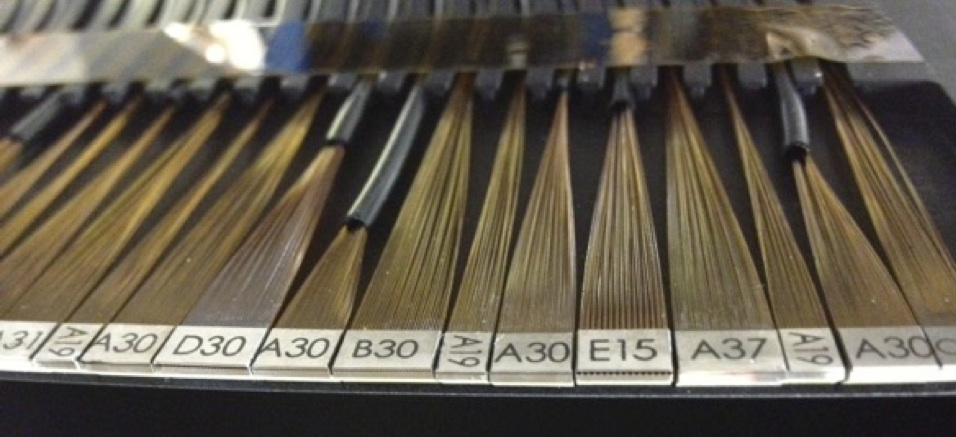}
\caption{V-groove blocks aligned onto the SP1 pseudo-slit for the P-MaNGA prototype instrument.  The V-groove blocks position the
  output end of the fibers that compose the IFUs and direct their light into the spectrographs.}
\label{fig:blocks}
\end{figure}

The bundle of optical fibers is wrapped in a protective cable and runs roughly 2 m from the input end of the IFU to the output end
which terminates in a V-groove block. As in the BOSS instrument, these blocks are a few millimeters in length and, when attached to each
spectrograph's pseudo-slit, serve to direct the light output from every fiber into the spectrographs. Bundled sets of 21--39
fibers fan out before reaching each block, and individual fibers are then laid in each groove, secured with a lid, and glued into
place (Figure \ref{fig:blocks}).  The final step is to polish the entire block flat so that the fibers and block surface can be
arranged on a curved slit plate to conform approximately to the entrance angle required by the two spectrographs.  As described in
N.~Drory et al.~(in preparation), an analysis of Hartmann exposures indicates that the fiber beam angle is correctly aligned at the 0.14 degree level.
The mapping between a fiber's 2D location within each IFU and its 1D location on a V-groove block is chosen to minimize the chance
of having fibers with significantly different flux adjacent to one another, thus limiting the impact of cross talk between fiber
traces on the CCD. As with the IFU end, the termination of fibers in their V-groove blocks is performed by CTechnologies.

An important consideration for MaNGA has been the acceptable ``slit density'' or number of V-grooves per unit length. In BOSS, the
fiber-to-fiber V-groove block spacing was 266~$\mu$m.  This produces spots on the CCD of width $\sim$40~$\mu$m separated by
101~$\mu$m. The nearly Gaussian spots minimally overlap in the ``spatial'' direction, resulting in what we refer to as
``fiber-trace cross talk'' of $\approx$1\%. In MaNGA, adjacent fibers on the sky experience a kind of cross talk from the
atmospheric seeing that dominates over the fiber-trace term. Our simulations show that for typical atmospheric conditions, we
expect $\approx$6\% seeing-induced cross talk between fiber pairs. Since each fiber interior to an IFU has six neighbors on-sky, but
only two along the slit, we can afford to pack the fibers significantly closer before the fiber-trace cross talk becomes significant.
One of the major goals of the MaNGA prototype instrument was to test the maximum slit density that could be reliably extracted and
tolerated given the science requirements.  Three fiber spacings were tested, and the tightest spacing of 177~$\mu$m (for the
science IFUs) was ultimately adopted for the final instrument, resulting in $\sim$10\% cross talk between the spectral traces of
adjacent fibers.  This cross talk can be effectively subtracted by forward-modeling each fiber's spatial profile using an
optimal-extraction method (see \citealt{bolton12}, D.~Schlegel et al.~(in preperation), D.~R.~Law et al.~(in preparation)).  The spacing on V-groove blocks
associated with mini-bundles is 204~$\mu$m.

For each MaNGA cartridge (see below), the pseudo-slits for both spectrographs contain 22 V-groove blocks each. These blocks
represent the output of 17 science IFUs, 12 seven-fiber mini-bundles for standard stars, and 92 sky fibers (see Table
\ref{tab:hardware}).  The total number of fibers is 1423.

\subsection{The BOSS Spectrographs}

MaNGA makes use of the two identical, dual-beam SDSS/BOSS spectrographs, each of which features a red and blue arm for continuous
wavelength coverage from the near-UV into the near-IR.  Details on the spectrographs are given in \citet{smee13}.

While similar in design to the original SDSS spectrographs, a 2009 upgrade as part of SDSS-III resulted in nearly a factor of two
improvement in the peak sensitivity, a wider wavelength coverage, and an increase in the number of accommodated fibers compared
to the original SDSS.  The spectrographs are secured to the back of the telescope on either side of the cartridge bay so that when cartridges are mounted into place, the fiber output as guided by the cartridge's two slitheads aligns with the input of the corresponding spectrograph.  

After reflecting off of a shared collimator, the input beam in each spectrograph is split into blue and red channels at 605 nm.
Each channel features its own camera optics, CCD, and VPH grism.  An index modulation density of 400 lines mm\per\ is used in the red
channel VPH grating and 520 lines mm\per\ in the blue channel.  The peak grating efficiencies are 82\% and 80\%, respectively. The
resulting spectral resolution rises on the blue side from $R \sim 1400$ at 4000 \AA\ to $R \sim 2100$ at 6000 \AA.  On the red
side, $R \sim 1800$ just beyond 6000 \AA\ and peaks at $R \sim 2600$ near 9000 \AA.

Both the E2V blue-channel CCD and the red-channel LBNL fully depleted CCD have a 4k $\times$ 4k format with 15 $\mu$m pixels.  The
read noise is less than 3.0 e$^{-}$ pixel\per\ rms (blue) and 5.0 e$^{-}$ pixel\per\ rms (red) and the dark current is $\sim$1 e$^{-}$ per
15-minute exposure.  Each fiber projects a spectrum onto the CCDs with a spatial FWHM of $\sim$2.3 pixels and a
line-spread function of width $\sim$2.3 pixels.  The full-frame readout time is 70 s.

The total system telescope$+$fiber$+$spectrograph throughput is $\sim$25\% over most of the bandpass (4500--9000 \AA) and greater than 10\% at all wavelengths from 4000 to 10,000 \AA.

\subsection{Plug-plate Cartridge System}\label{sec:facility}

Spectroscopic observations using the Sloan Telescope are performed with a sophisticated plug-plate cartridge system.  The
cartridges have the shape of a thick disk, measuring roughly 1 m in diameter and 60 cm in height, with two shoebox-like housings
mounted on either side that cover the delicate slitheads.  The cartridges are composed of a metal frame, with internal anchoring
hardware that holds both the fiber harnesses and the plug-plate, and an outer plastic casing that can be removed to plug the
fibers \citep[for details, see][]{smee13}.

Aluminum SDSS plug-plates are drilled for specific fields and target sets at the University of Washington and shipped to Apache Point Observatory (APO) several months in advance of observation.
During the day, SDSS employs fiber optic technicians who mount plates in available cartridges and plug them.  Weighing roughly 140
kg, these cartridges are hydraulically installed onto the telescope and exchanged over the course of observations throughout the night.

At the end of SDSS-III, cartridges were assigned to either BOSS or APOGEE, which determined the type and number of fibers
installed.  Because APOGEE uses 300 fibers to interface with a near-IR bench-mounted spectrograph in a separate building, MaNGA
fibers and IFUs can be plugged simultaneously with APOGEE fibers, allowing both instruments to observe at the same time.  MaNGA
components will be installed into six cartridges in SDSS-IV, yielding a total of 102 science IFUs in operation at any given time.

\subsection{Spectrophotometric Calibration}\label{sec:calibration}

The MaNGA survey requirement on spectrophotometry is a relative accuracy\footnote{Absolute calibration of MaNGA data can be
  accomplished by referencing broadband SDSS imaging.} of better than 7\% from \ion{O}{2}~$\lambda3727$ to H$\alpha$ and better than
2.4\% between H$\beta$ and H$\alpha$. This precision ensures that calibration errors do not dominate the error budget on galaxy SFRs and
nebular metallicities. However, the implementation of flux calibration for IFUs is different than that for single fibers or slits
because with an IFU, there is no need to correct for the light lost from a fixed aperture. The beam at different wavelengths---as a
result of the wavelength-dependent point-spread function (PSF) and DAR---is simply shifted to different portions of
the IFU.

The goal is to measure the flux loss purely due to the atmosphere and system throughput, without losses caused by aperture effects
or DAR. In the prototype run, we experimented with a number of options including larger fiber sizes to capture more light for
standards. In the final survey, we employ 12 seven-fiber ``mini-bundles'' when targeting spectrophotometric standards. Our tests
have verified that when combined with an initial estimate of the seeing PSF from the guide camera, the seven-fiber flux
measurements provide enough information for a detailed fit of the fraction of light covered by the central fiber as a
function of wavelength, thus enabling a measurement of the flux loss due to the system throughput alone.  Further details on the
strategy, implementation, and performance of spectrophotometric flux calibration are presented in R.~Yan et al.~(in preparation).

\subsection{Sky Subtraction}\label{sec:sky}

Sky subtraction represents a second calibration challenge that motivated an intensive effort for the final MaNGA design and was a
major topic of study in the prototype data analysis. This topic is covered in detail in the instrument description paper (N.~Drory et
al., in preparation) and the software pipeline paper (D.~R.~Law et al., in preparation) as sky subtraction performance involves an
interplay between hardware design and software algorithms.  The BOSS reduction pipeline \citep{bolton12}, on which the MaNGA
reduction software is based, was able to achieve $\sim$1\% level sky subtraction in the continuum. 

MaNGA's goal is to achieve Poisson-limited subtraction in the core of bright skylines even at the reddest wavelengths with a
tractable number of fibers that does not degrade survey efficiency. We have chosen a hardware solution that locates sky fibers
near science IFUs on the plate and ensures that associated sky and science fibers are also co-located along the slit.  We add two
sky fibers onto every V-groove block that composes a science IFU\footnote{Each mini-bundle is assigned one sky fiber.  The
  V-groove blocks associated with mini-bundles group three sets of mini-bundle fibers together; 21 fibers for standard star
  calibration and three sky fibers.}.  The sky fibers are placed on the ends of the blocks and their adjacent science fibers are
required to be in the periphery of the IFU so as to minimize contamination from bright sources.  The sky fibers can patrol a sky
region within 14\arcmin\ of each IFU.  The number of associated sky fibers ranges from 2 for the 19-fiber bundles to 8 for the
127-fiber bundles.  A total of 92 sky fibers are available in each cartridge, 46 spread uniformly along each spectrograph
pseudo-slit.

\begin{deluxetable}{lcc}
\tablecolumns{3}
\tablewidth{0pc}
\tablecaption{Hardware Summary\label{tab:hardware}}
\tabletypesize{\footnotesize}
\tablehead{
\colhead{Component}  & \colhead{P-MaNGA} & \colhead{MaNGA} \\
\colhead{}                    &\colhead{(Prototype)}  &\colhead{(Survey)} \\
}
\startdata

Number of carts & $\sfrac{1}{2}$ & 6 \\
Spectrographs used & sp1 & sp1, sp2 \\
N$_{19}$ (12\arcsec\ diam.) & 5 & 2 \\
N$_{37}$ (17\arcsec\ diam.)  & 0 & 4 \\
N$_{61}$ (22\arcsec\ diam.) & 1 & 4 \\
N$_{91}$ (27\arcsec\ diam.) & 0 & 2 \\
N$_{127}$ (32\arcsec\ diam.) & 2 & 5 \\
N$_{7}$ (mini-bundles) & 0 & 12 \\
N$_{\rm sky}$ per spec. & 41 & 46 \\
Sky fiber type & Plate-roaming, & IFU-proximate, \\
& separate block & plate and slit \\
N$_{\rm standard}$ per spec. & 6 & 6 \\
Flux standard type & Single 2\arcsec\ fiber & Mini-bundle \\
V-groove pitch (IFUs) & 204$\mu$m & 177$\mu$m \\ 
Additional fibers & 60$\times$3\arcsec; 30$\times$5\arcsec & none \\
Total fibers per cart & 560 & 1423 \\

\enddata
\tablecomments{IFU sizes are indicated by N$_{X}$ where $X$ is the number of fibers per IFU.  The number of IFUs of a given size are listed per cart.}
\end{deluxetable}

\subsection{P-MaNGA: Prototype Hardware Summary}\label{sec:p-manga-hardware}

The P-MaNGA prototype instrument was installed in 2012 December and used to conduct a series of tests and observations (see
\S\ref{sec:obs}) in 2012 December and 2013 January.  As shown in Table \ref{tab:hardware}, the P-MaNGA prototype differed in many
respects from MaNGA's final survey instrumentation.  First, P-MaNGA used only 560 total fibers distributed across just one of the
two BOSS spectrographs.  A total of 470 of these fibers had a standard size of 2\farcs0 core diameter, and 60 of these were used
for sky subtraction.  To experiment with flux calibration options that could potentially correct for aperture losses on standard
star observations, we deployed an additional 90 fibers with either a 3\farcs0 or 5\farcs0 core diameter.  Our tests with
these fibers showed poorer results compared with experiments in which we placed the P-MaNGA IFUs on standard stars and examined only
the flux from the central seven fibers to simulate mini-bundle observations. These experiments showed that 12 mini-bundles would deliver
the statistical precision required for the flux calibration vector with the minimum use of fiber ``real estate'' along the slit.

The remaining 410 2\arcsec\ fibers were bundled into eight IFUs with three sizes: 19 fibers (\N{19}), 61 fibers (\N{61}), and 127 fibers (\N{127}).  The P-MaNGA IFU complement (Table \ref{tab:p-manga-ifu}) was dramatically different from the MaNGA survey instrument, with $5 \times $\N{19} (instead of just 2), $1 \times $\N{61} (instead of 4), $2 \times $\N{127} (instead of 5), and no 37-fiber or 91-fiber IFUs.

In P-MaNGA, sky fibers were entirely allocated to specific V-groove blocks and plugged in random positions across the full field
of view. The MaNGA instrument will colocate sky fibers near IFU targets both on the sky and on the slit with the
goal of tracking apparent sky (and specifically skyline) variations both across the field of view and as a result of subtle
differences in the line-spread function, scattered light, and spectral extraction along the spatial dimension of the spectroscopic
data.

As discussed in \S\ref{sec:blocks}, a key question addressed by the prototype instrument was the optimum ``slit density'' or
fiber-to-fiber pitch on entrance to the spectrographs.  P-MaNGA bundle {\tt ma003} was a particularly important test case.
Dubbed the ``Frankenbundle,'' its 127 fibers were fed into four V-groove blocks, one with a pitch of 260 $\mu$m, two
with a pitch of 204 $\mu$m, and one with a pitch of 177 $\mu$m.  Three key observational tests were performed with the
Frankenbundle to conclude that the 177 $\mu$m pitch was acceptable, and this value was adopted for all science IFUs in the final
MaNGA instrument.  First, a trio of closely separated stars was observed with this IFU at a specific orientation so that each star
landed on portions of the IFU corresponding to a different V-groove slit density.  Under simultaneous seeing conditions, this
allowed us to measure the fiber trace cross talk and confirm that it was acceptable (less than $\sim$10\%) for the tightest spacing.
Second, several galaxies were observed with the Frankenbundle, and maps of H$\alpha$ and stellar velocity fields as well as flux
maps in both line emission and the continuum showed no patterns matching the arrangement of fibers with different slit densities
in the IFU.  Third, the same galaxy was observed with both the Frankenbundle and the {\tt ma008} IFU, which had a uniform slit
density of 204 $\mu$m.  No significant differences in observed quantities were detected.  Finally, in addition to these observational tests, a
number of simulated observations exploring different levels of cross talk demonstrated no detectable differences up to 15\%
fiber-trace cross talk and only minimal effects up to as high as 25\%.  With the final choice of 177 $\mu$m, MaNGA is able to make
use of 1423 fibers fed to the same spectrographs that in the BOSS instrument only accommodate 1000 fibers.

Another important difference with P-MaNGA data is a degradation in sky subtraction owing to a problem in the BOSS SP1 spectrograph
during the P-MaNGA observations.  Especially redward of $\sim$8000 \AA, a significant coma aberration in the red channel of the SP1
spectrograph was discovered by the MaNGA team, which severely impacts the modeling and subtraction of bright skylines at red
wavelengths, and increasingly for fibers near the slit edges.  We have traced its appearance to an upgrade of various spectrograph
components that was performed in 2011 August.  This issue was not critical for the successful completion of the BOSS survey;
SDSS-IV has now fixed the coma problem.

\begin{deluxetable}{lcccc}
\tablecolumns{5}
\tablewidth{0pc}
\tablecaption{P-MaNGA IFUs \label{tab:p-manga-ifu}}
\tabletypesize{\footnotesize}
\tablehead{
\colhead{Bundle}  & \colhead{Name} & \colhead{N$_{\rm fibers}$} &\colhead{N$_{\rm radial}$} &\colhead{$\Theta_{\rm diameter}$} \\
\colhead{} &\colhead{}  &\colhead{} &\colhead{} &\colhead{(arcsec)} \\
}
\startdata
ma001 & 19E & 19 & 3 & 12.5 \\ 
ma002 &  61A & 61 & 5 & 22.5  \\ 
ma003$^{\dagger}$ & 127A & 127 & 7 & 32.5 \\ 
ma004 & 19A & 19 & 3 & 12.5 \\ 
ma005 & 19B & 19 & 3 & 12.5 \\ 
ma006 & 19C & 19 & 3 & 12.5 \\ 
ma007 &  19D & 19 & 3 & 12.5 \\ 
ma008 &  127B & 127 & 7 & 32.5
\enddata
\tablecomments{N$_{\rm fibers}$ gives the number of fibers in the IFU.  N$_{\rm radial}$ indicates the corresponding number of hexagonal rings, including the central fiber.  \\ $^{\dagger}$ Bundle ma003 (127A), a.k.a.\ ``Frankenbundle,'' employed V-groove blocks with multiple fiber spacings.}
\end{deluxetable}

\medskip

\section{Sample Design Concept}\label{sec:survey}\label{sec:sample}

An intensive effort to design and optimize the MaNGA target selection under various constraints and requirements is presented in
D.~A.~Wake et al.~(in preparation), who also give details on the final selection cuts applied and the resulting sample sizes and their
distributions in various properties.  In what follows, we present a brief conceptual summary based on a near-final set of
selection criteria.

The MaNGA sample design is predicated on three major concepts. First we require a sample of roughly 10,000 galaxies, the
motivation of which is discussed in \S\ref{sec:discovery}.  Second, we seek a sample that is complete above a given stellar mass
limit, for which we adopt $\log M_*/\Msun > 9$, and has a roughly flat $\log M_*$ distribution. This selection is motivated by ensuring
adequate sampling of $M_*$, widely recognized as a key parameter or ``principal component'' that defines the galaxy population.
Third, we choose to prioritize uniform radial coverage in terms of the scale length associated with the galaxy's light profile,
which we quantify using the major-axis half-light radius, \Reff. This choice acknowledges the self-similar behavior of many
aspects of galaxy structure and ensures a consistent picture of the outer versus inner regions of sample galaxies as defined by
their surface brightness profiles.  

With these basic tenets adopted, we then optimize our design to maximize the spatial resolution and the S/N obtained per unit
physical area, both of which favor targets at the lowest redshifts possible. The need for sufficient volume to build the sample
sizes required, coupled with the goal of covering intrinsically larger galaxies to the same radius in terms of \Reff, favors
targets at higher redshifts.  

\begin{figure}
\centering
\includegraphics[scale=0.55, angle=0]{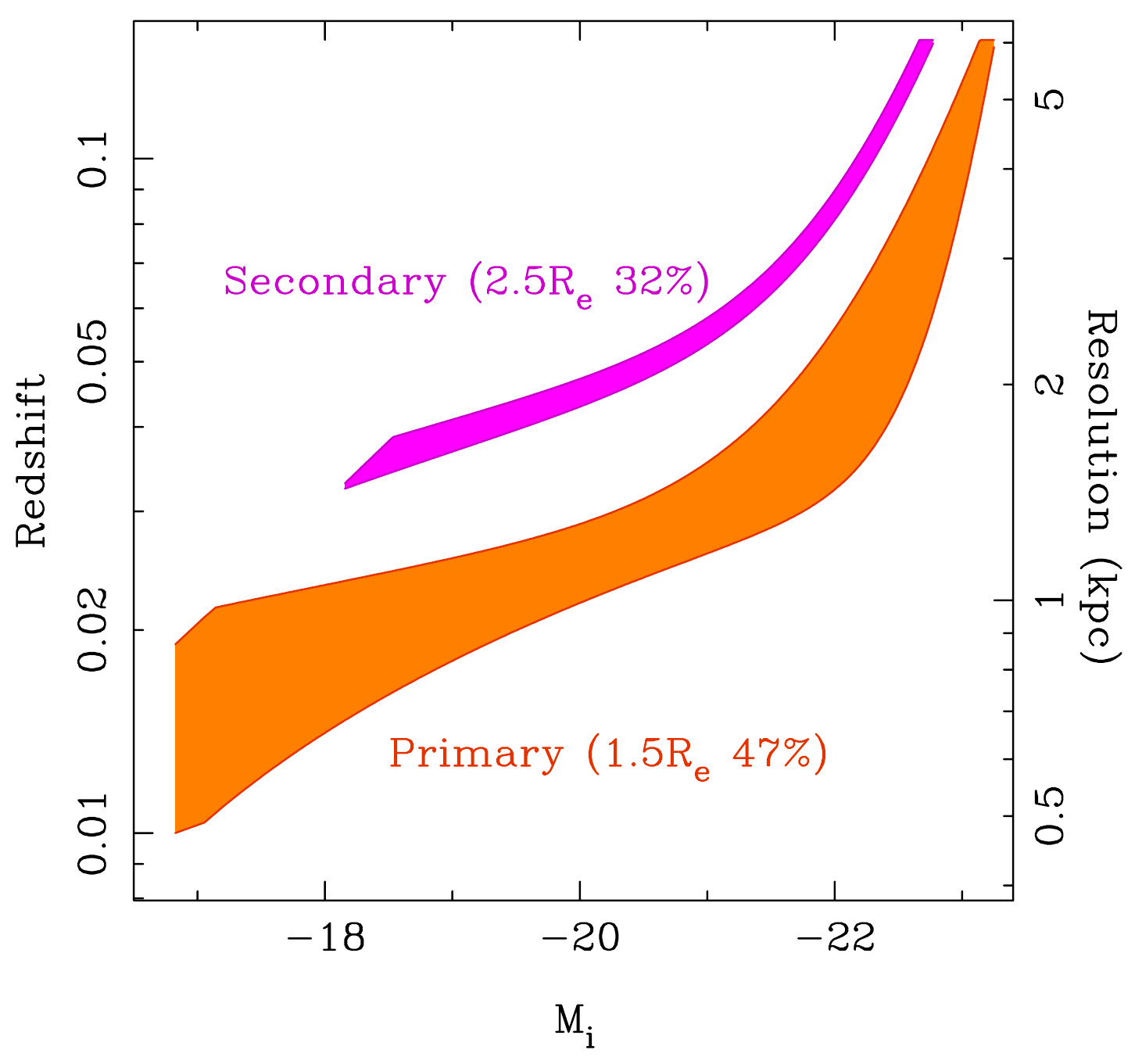}
\caption{Preliminary redshift selection cuts defining the MaNGA Primary and Secondary samples (v1.beta).  All galaxies within
  the redshift boundaries are targeted, resulting in volume-limited samples defined as a smooth function of $i$-band magnitude.
  The nature of the selection is designed to maintain uniform coverage to 1.5 \Reff\ for the Primary sample and 2.5 \Reff\ for the
  Secondary sample while providing a roughly flat distribution in $M_*$.  The corresponding spatial resolution assuming an
  effective reconstructed PSF with FWHM$ \sim $2\farcs5 is indicated on the right-hand axis.  The percentages indicated assume
  that 10\% of MaNGA targets are reserved for ancillary sources.}
\label{fig:sample_cuts}
\end{figure}

Throughout this process, we select almost entirely on SDSS Main Galaxy Sample\footnote{We select using an extension of the
  NASA-Sloan Atlas (NSA; Blanton M. {\tt http://www.nsatlas.org}). The NSA is a catalog of nearby galaxies within 200 Mpc ($z =
  0.055$) based on the SDSS DR7 MAIN galaxy sample \citep{abazajian09}, but incorporating data from additional sources and image
  reprocessing optimized for large galaxies \citep{blanton11}.  For MaNGA, the NSA has been extended to $z = 0.15$.} absolute
magnitudes and redshifts \citep{strauss02}, thereby avoiding ``black boxes'' (such as specific code used to estimate $M_*$ or
SFR).  We use these quantities to define a luminosity-dependent volume-limited sample, as shown by the shaded redshift intervals
in Figure \ref{fig:sample_cuts}.  The rise in these intervals with increasing luminosity ensures that more luminous, intrinsically
larger galaxies are targeted at greater distances so that they can be covered to the same radius and so that sufficient numbers
can be obtained to balance the final $M_*$ distribution.  This results, however, in a decrease in the physical resolution obtained
at brighter luminosities (right-hand axis in Figure \ref{fig:sample_cuts}).

We select ``Primary'' and ``Secondary'' samples defined by two radial coverage goals.  The Primary selection ($\langle z \rangle =
0.03$) reaches 1.5 \Reff\ (for more than 80\% of its targets) sampled with an average of five radial bins.  It accounts for
$\sim$5000 galaxies.  To the Primary selection we add a ``Color-Enhanced sample'' of an additional $\sim$1700 galaxies designed
to balance the color distribution at fixed $M_*$.  The color enhancement increases the number of high-mass blue galaxies, low-mass
red galaxies, and ``green valley'' galaxies tracing important but rare phases of galaxy evolution.  We refer to these two
selections as ``Primary+''.

The Secondary selection of $\sim$3300 galaxies ($\langle z \rangle = 0.045$) is defined in an identical way to the Primary sample,
but with a requirement that 80\% of the galaxies be covered to 2.5 \Reff. Naturally this criterion results in higher-redshift
selection limits and lower spatial resolution than the Primary sample (Figure \ref{fig:sample_cuts}). Together, the Primary+ and
Secondary samples represent $\sim$90\% of all MaNGA targets.  The remaining $\sim$10\% will be dedicated to ancillary targets of
high value outside the main selection, including a subset of low-$z$ luminous galaxies that will provide $\sim$1 kpc resolution, albeit with
less radial coverage.  An initial set of ancillary targets will be chosen within the first several months of the survey.  With
assumptions on weather conditions and overhead, and applying the MaNGA observing limits to the results of survey tiling and IFU
allocation exercises, our survey simulations indicate that MaNGA will observe 10,400 galaxies over its 6 yr
lifetime.

\subsection{Survey Footprint}

MaNGA plates can be targeted over the full footprint of the SDSS Main spectroscopic sample and in total will cover $\sim$2700
deg$^2$. In practice, during the fall observing season, this means that the MaNGA footprint is relegated to the equator
\citep[Stripe 82, see][]{stoughton02} and one other stripe in the South Galactic Cap (SGC). While SDSS-III imaged more of the SGC,
the spectroscopic redshifts required of MaNGA targets (with declinations above zero) are limited to two narrow stripes. This is
not the case in the North Galactic Cap (NGC, spring observing) where the full extragalactic sky is available. The higher average
airmass and extinction of equatorial fields result in an efficiency loss of nearly 40\%. Therefore, from the point of view of
efficiency alone, MaNGA prefers mid-declination fields in the NGC but has no choice but to observe at the equator in the SGC. The
ideal field choice on any given night is modulated by the airmass and declination limits set by our requirements on the time
variation of DAR. Other considerations on field choice will weigh the value of ancillary survey overlap (especially with radio
surveys and deep wide-field imaging surveys) and will be decided in the early stages of SDSS-IV.

\begin{figure}
\centering
\includegraphics[scale=1.45, angle=0]{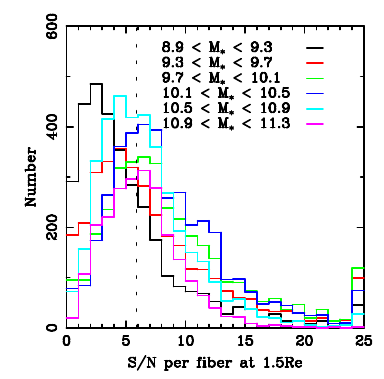}
\includegraphics[scale=1.45, angle=0]{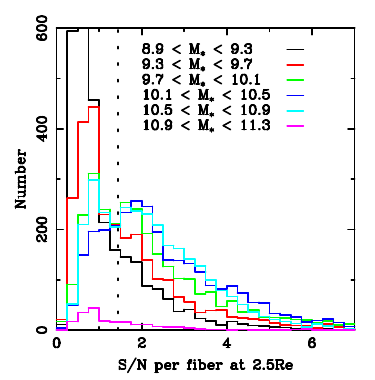}
\caption{For a realization of the MaNGA sample, representative distributions in the per-fiber $r$-band S/N per pixel obtained for
  MaNGA's Primary (top) and Secondary sample (bottom) for target galaxies in different $M_*$ bins, assuming a nominal 3 hr total
  exposure time in median conditions. S/N estimates are based on actual SDSS surface brightness profiles.}
\label{fig:sn_dist}
\end{figure}

\subsection{Observing Strategy}\label{sec:obs}

We provide a brief summary of MaNGA's observing strategy.  Full details are described in D.~R.~Law et al.~(in preparation) and R.~Yan et al.~(in preparation).

MaNGA observations of a field are performed in sets of three exposures, with a sub-fiber-diameter offset dither applied between
exposures. The integration time of 15 minutes is chosen to minimize variation in the parallactic angle during exposures that would
lead to unacceptable shifts in the fiber exposure map at the extremes of the wavelength range. On average, three to four sets (sometimes
obtained on different nights) will be required before a plate will be deemed complete based on the achieved total S/N as
determined in near real-time. The S/N goals are set by the science requirements and correspond to 3 hr total exposure times in
median conditions. The expected S/N distributions for the two main samples are presented in Figure \ref{fig:sn_dist}.

Although the observational seeing is expected to be $\sim$1\farcs5, the reconstructed PSF in combined datacubes after dithering
and fiber sampling is $\sim$2\farcs5 (FWHM).  Because of the regular hexagonal packing of MaNGA IFUs, a defined three-point dither
pattern can be adopted that achieves uniform spatial sampling for all targets.  To ensure that the resulting exposure map provides even
coverage at all wavelengths in the face of DAR, observations are restricted to observability windows set by the field declination
that ensure acceptable levels of variance over the course of an hour-long dither set.  Once a three-exposure set is initiated, it
should be completed (possibly on subsequent nights) at the same corresponding hour angle.  Different sets initiated at different
times within the observability window are combined in the reduction pipeline with acceptance criteria based on the atmospheric
seeing (FWHM$ < 2$--2.5\arcsec) and transparency.

\begin{figure*}[ht!]
\centering
\epsscale{1.1}
\plotone{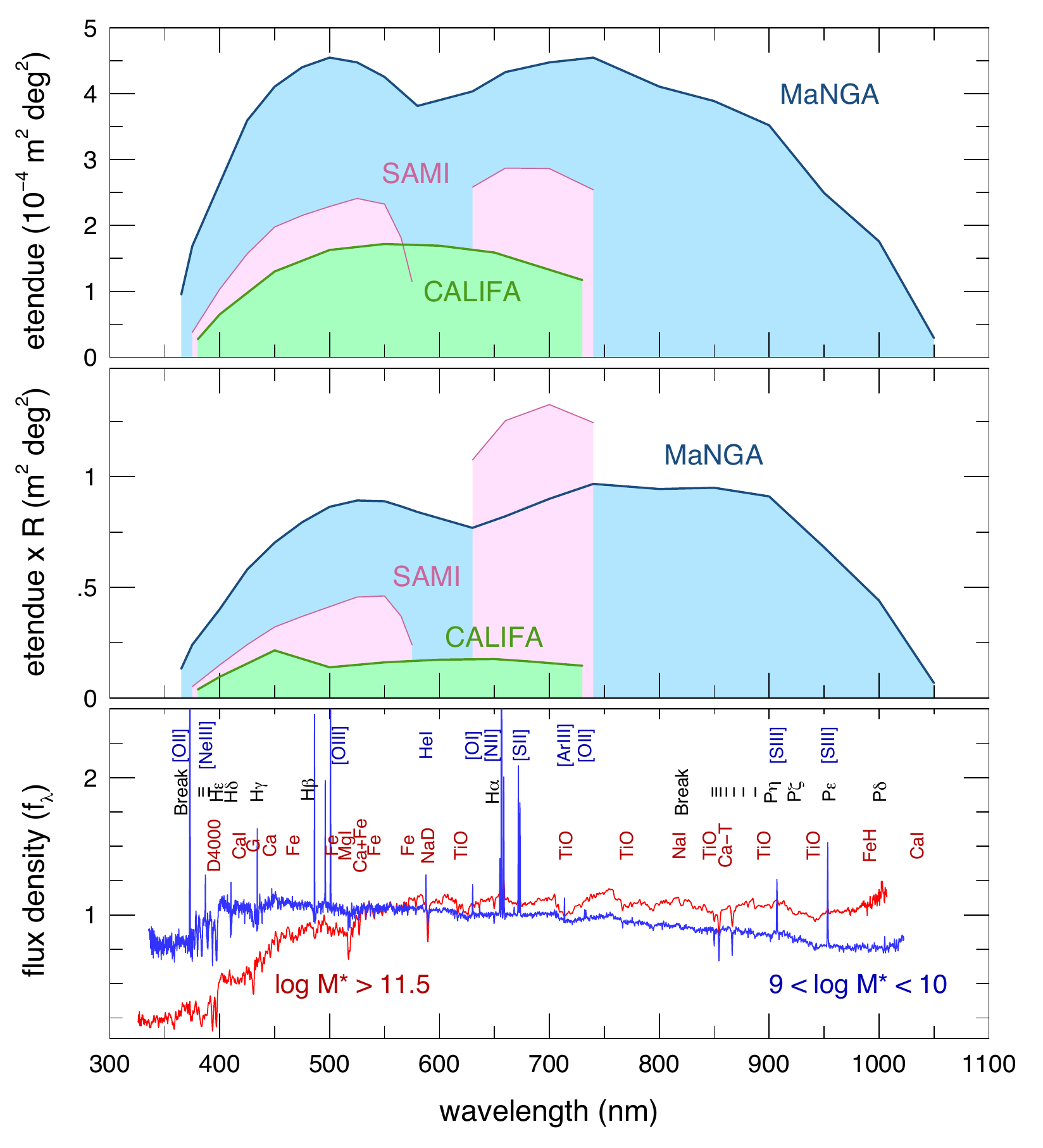}
\caption{MaNGA's simultaneous information gathering power, reckoned
by two metrics, \'etendue (top panel), and $R \times $\'etendue (middle panel), as
defined in the text. Two examples of stacked $z \sim 0.15$ spectra
from the BOSS survey \citep[bottom;][]{dawson13} illustrate the
available spectral features for ISM and stellar
composition and kinematic analysis in galaxies bracketing the sample
range in $M_*$.}
\label{fig:etendue}
\end{figure*}

%
\begin{deluxetable*}{lcccccc}
\tablecolumns{7}
\tablewidth{0pc}
\tablecaption{Comparison of IFU Surveys\label{tab:surveys}}
\tabletypesize{\scriptsize}
\tablehead{
\colhead{Specification}  & \colhead{MaNGA} & \colhead{SAMI}  & \colhead{CALIFA} & \colhead{\specialcell[t]{DiskMass \\ (H$\alpha$)}} & \colhead{\specialcell[t]{DiskMass \\ (stellar)}} & \colhead{\atl} \\
}
\startdata

Sample size & 10,000 & 3,400 & 600 & 146 & 46 & 260 \\
\\[-5pt]
Selection & $M_* > 10^9$\msun & $M_* > 10^{8.2}$\msun & 45\arcsec$ < D_{25} < 80 $\arcsec & \multicolumn{2}{c}{\specialcell{S/SAab-cd, b/a$>$0.75 \\ 10\arcsec$<$h$_{R}$$<$20\arcsec}} & \specialcell{$M_* \gtrsim 10^{9.8}$\msun$^e$ \\ E/S0}  \\
\\[-5pt]
Redshift & 0.01--0.15 &  0.004--0.095  & 0.005--0.03 & 0.001--0.047  & 0.003-0.042 & $z \lesssim 0.01$ \\
\\[-5pt]
Radial coverage & \specialcell{1.5 \Reff (P+)\\2.5 \Reff (S)} & 1.1--2.9 \Reff & 1.8--3.7 \Reff & 1.4--3 \Reff  & 1.1--2.3 \Reff & 0.6--1.5 \Reff \\
\\[-5pt]
\lspecialcell{S/N$^a$ at 1\Reff \\ (per spatial sample)} & 14--35 & 12--28 & 10--50 & 6 & 9--16 & 15 \\
\\[-5pt]
$\lambda$ range (nm) & 360--1030 & \specialcell{370--570 (580V)\\625--735 (1000R)} & \specialcell{375--750 (V500)\\370--475 (V1200)} & 648--689 & 498--538 & 480--538 \\
\\[-5pt]
$\sigma_{\rm instrument}$ (km s$^{-1}$) & 50--80 &  \specialcell{75\\28} &  \specialcell{85 \\150 } & 13 & 16 & 98 \\
\\[-5pt]
\lspecialcell{Angular sampling$^b$\\ (diameter)} & 2\arcsec & 1\farcs6 & 2\farcs7 & 4\farcs7 & 2\farcs7 & 0\farcs8 \\
\\[-5pt]
\lspecialcell{Angular FWHM \\ (reconstructed)} & 2\farcs5 & 2\farcs1$^c$ & 2\farcs5 & 6\arcsec & 3\farcs5 & 1\farcs5 \\
\\[-5pt]
\lspecialcell{Spatial FWHM \\ (physical)} & \specialcell{1.3--4.5 kpc (P+)\\2.2--5.1 kpc (S)} & 1.1--2.3 kpc & 0.8--1.0 kpc & 0.4--4.2 kpc & 0.3--3.0 kpc & 0.15 kpc\\
\\[-5pt]
\lspecialcell{Spatial FWHM \\ (in \Reff)} & \specialcell{0.2--0.6 (P+)\\0.3--0.9 (S)} & 0.3--0.8 & 0.2 &0.2--0.4 & 0.1--0.2 & 0.09 \\
\\[-5pt]
IFU fill factor & 56\% & 73\% & 53\% & 25\% & 53\% & 100\% \\
\\[-5pt]
\lspecialcell{With gradients \\ measurable$^d$ to} & & & & & & \\
\hspace{0.5cm} 1.0 \Reff :  & 4070 &  720 &  580 & 128 & 39 & 112 \\
\hspace{0.5cm} 1.5 \Reff :  & 6050 &  790 &  521 & 122 & 20 & 47 \\
\hspace{0.5cm} 2.0 \Reff :  & 2570 &  680 &  462 & 80 & 5 & 26 \\
\hspace{0.5cm} 2.5 \Reff :  & 2340 &  460 &  340 & 26 & 0 & 13 \\
\hspace{0.5cm} 3.0 \Reff :  & 670 &  350 &  111 & 3 & 0 & 1 \\

\enddata
\tablecomments{When ranges are given (other than redshift and wavelength), we use 20th and 80th percentile estimates.  $D_{25}$ is related to
  the SDSS ``isoA\_r'' major-axis diameter of the 25 AB arcsec$^{-2}$ $r$-band isophote.   For MaNGA and SAMI, radii are defined in terms of \Reff\ as
  measured for \sersic\ fits performed in the NASA Sloan Atlas.  For CALIFA, half-light radii were remeasured from the imaging data
  by the CALIFA team, and for DiskMass, \Reff\ are estimated from measured disk radial scale lengths (h$_R$) adopting
  \Reff/h$_R$$\sim$1.67 \citep{lackner12a}.  For MaNGA, ``P+'' refers to the Primary+ Sample which accounts for $\sim$\sfrac{2}{3}
  of the survey targets.  The Secondary sample ($\sim$\sfrac{1}{3}) is designated by ``S.'' \\ 
$^a$ S/N is given per $\lambda$ resolution element, per spatial element (e.g., per fiber), in the $r$-band. \\ 
$^b$ The angular sampling diameter is either that of a fiber or the width of the lenslet in the case of \atl. \\ 
$^c$ The expected SAMI FWHM averaged over the full survey is 2\farcs1 (S.~Croom, private communication).  Results from the first
year indicate a value of 2\farcs4 \citep{sharp15}.  \\
$^d$ A target galaxy is defined to have a ``measurable gradient'' within the specified radius if the IFU field-of-view covers this
radius with more than 2.5 spatial resolution elements with size given by the reconstructed FWHM.  In the case of MaNGA, the same
galaxy may be included in multiple bins (i.e., the total number of measurable gradients is $\sim$1.5 times the final sample size).
For SAMI, the bins are unique. \\
$^e$ This is only approximate.  \atl\ selected visually-classified early-type galaxies with $M_K < -21.5$, $D < 42 $Mpc, $|{\rm Dec} - 29^{\circ}| < 35^{\circ}$, and $|b| > 15^{\circ}$.
}
\end{deluxetable*}

\section{IFU Survey Comparison}\label{sec:comparison}

MaNGA follows in the footsteps of previous IFU surveys, not only in increasing sample size, but in terms of building on important
lessons learned regarding design, instrumentation, and analysis. These surveys were introduced in \S\ref{sec:intro}.  In this
section, we describe a more detailed comparison of instrumentation and survey designs, focusing on campaigns targeting more than
$\sim$100 galaxies: \atl, DiskMass, CALIFA, SAMI, and MaNGA.

In Figure \ref{fig:etendue}, we compare two metrics of the simultaneous information-gathering power of the CALIFA, SAMI, and MaNGA
survey instrumentation, displayed as a function of wavelength. Relevant to both metrics is the total system efficiency
($\epsilon$) which is adopted from the literature for CALIFA \citep{roth04, kelz06, sanchez12} and SAMI \citep{bryant14}. For
MaNGA, $\epsilon$ is a factor of 1.1 greater than the BOSS throughput curves from \citet{smee13}, who did not correct for the PSF
aperture losses from a 2\arcsec\ diameter fiber in FWHM 1\farcs1 seeing. In this way, our definition of $\epsilon$ can be
described as the ratio of the flux detected from a source with uniform surface brightness divided by the incident flux from this
source (before atmospheric losses) that could have been collected by the IFU\footnote{In the case of SAMI, \citet{bryant14} report
  throughput curves for the first generation SAMI instrument that are not consistent with those measured for the same
  instrumentation in \citet{Croom2012}.  This
  reflects updates to the throughput calculation performed in \citet{bryant14}, including the adoption of the same definition of
  $\epsilon$ used here (Bryant, J., private communication).}.

In the top panel of Figure \ref{fig:etendue}, \'etendue is the product of the telescope collecting area ($A$; accounting for the central
obstruction), solid angle covered by all fibers ($\Omega$), and $\epsilon$. In the photon-limited regime, \'etendue is a measure
of how quickly one can map the sky to a given S/N at a given spectral resolution. While MaNGA's grasp ($A \times
\Omega$) is slightly lower than that of CALIFA or SAMI (respectively 1.27, 1.30, and 1.41 in units of $10^{-3}$ m$^2$ deg$^2$), by
virtue of its system efficiency, MaNGA's \'etendue is greater and spans a broader range of wavelength.

What \'etendue does not capture is the trade-off between spectral resolution and wavelength coverage. For instrumental velocity
resolutions of $\sigma_{\rm instrument} > 5$--10 km s$^{-1}$ (resolving powers of $R \sim 13,000$ to $R \sim 25,000$), which
represent the coldest dynamical scales of unresolved structure in galaxies, any increase in spectral resolution adds kinematic and
spectrophotometric information at fixed S/N. To account for this, in the middle panel we plot the product of \'etendue and the
resolving power ($R$). The area under these curves is invariant to spectral resolution and can therefore be considered a metric of
simultaneous survey information-gathering power.

We provide further quantitative details on both the instrumentation and associated survey designs in Table \ref{tab:surveys}. This
set of parameters---physical resolution, radial coverage and ability to measure gradients, S/N, spectral resolution,
wavelength coverage, and sample size and selection---is motivated by the science goals discussed in \S\ref{sec:science} and basic
design requirements described in \S\ref{sec:envelope}.

We consider \atl\ \citep{cappellari2011} in its own right and also as a representative of the earlier SAURON survey
\citep{dezeeuw2002}. Both employ the SAURON instrument with the observational setup described in \citet{bacon01}.  We note that
\atl\ additionally includes coordinated observations at radio and millimeter wavelengths.  Mounted on the William Herschel
Telescope, SAURON is a single, large-format IFU with a field of view of 33 $\times$ 41 arcsec$^2$ coupled to a lenslet array. The
primary goal of the IFU observations in SAURON and \atl\ was to measure stellar kinematic fields for early-type galaxies.  This is reflected in the sample
design, wavelength range, and radial coverage, although many other investigations beyond galaxy dynamics have been successfully
performed with these surveys.  Compared to MaNGA, CALIFA, and SAMI, the \atl\ sample has the highest spatial resolution, more than
order of magnitude greater than larger IFU surveys, although only somewhat higher than DiskMass. It also features the highest S/N
achieved per unit area. The limitations include a relatively small sample that is morphologically (visually) selected, limited
radial coverage, and a narrow wavelength range.

Turning to the DiskMass survey \citep{bershady10}, this program is a natural counterpart to \atl: a sample of similar size
but employing a design concept tuned to studies of disk galaxy dynamics.  As the name implies, DiskMass was motivated by the
specific goal of distinguishing the mass contributions of baryonic and dark matter components as a function of radius in
disk-dominated galaxies.  Predominantly face-on galaxies were targeted with instrument setups capable of achieving relatively high
velocity resolution ($\sim$14 km s$^{-1}$) in order to measure the out-of-plane, vertical velocity components of stars and gas in
nearby disks.  Compared with \atl, DiskMass reaches larger radii but with a spatial resolution that is somewhat more coarse though
still better than the average resolution achieved by the larger surveys (CALIFA, SAMI, and MaNGA).

The CALIFA survey makes use of the fiber-bundle PPak IFU coupled to the Potsdam Multi-Aperture Spectrograph on the 3.5 m
telescope at Calar Alto Observatory.  This is the same as for the DiskMass stellar survey, but with different spectrograph grating
configurations.  CALIFA aims to expand IFU coverage to galaxies of all type, increasing the sample size beyond \atl\ by roughly a
factor of three. An overview and initial assessment of the observations are presented in \citet{sanchez12} and extended in \citet{garcia-benito14} while \citet{walcher14}
present the sample selection.  CALIFA has also sought greater radial coverage ($r \sim 2$ \Reff), choosing a diameter selection
and a redshift range matched to the chosen wavelength setup that pushes their sample to distances approaching MaNGA and SAMI
($0.005 < z < 0.03$), resulting in lower spatial resolution compared to \atl.  As of this writing, CALIFA has completed 400 of its
600 targets.

The SAMI survey \citep{Croom2012} began in 2013 and shares more similarities with MaNGA than the other IFU surveys. Like MaNGA,
SAMI makes use of {\em multiple} fiber-bundle IFUs---12 dedicated to galaxy targets, with one placed on a standard star---across
its 1$^{\circ}$ diameter field of view. By the time it finishes in 2016, SAMI is expected to observe 3400 galaxies (they have
completed 1000 galaxies at the time of this writing). The sample design goals are similar in terms of $M_*$ and redshift to
MaNGA's, but there are important differences. First, SAMI IFUs are fixed in size (15\arcsec\ diameter), while MaNGA's range from
diameters of 12\arcsec\ to 32\arcsec.  Second, SAMI's IFU sky density is 15 deg$^{-2}$ compared to MaNGA's 2.4 deg$^{-2}$.
Whereas the MaNGA main samples are designed to reach specified radii (1.5 \Reff, or 2.5 \Reff) and sample these with a minimum
number of spatial elements (on average, 5), the SAMI sample spans wider distributions in radial coverage and spatial resolution.
SAMI also realizes a $\sim$30\% greater physical spatial resolution (i.e., kpc) for the central regions of galaxies with $\log
M_*/M_{\odot} > 10.5$ but does not always cover their larger sizes.  Aside from the wider wavelength coverage in MaNGA, another
major difference between SAMI and MaNGA is SAMI's greater spectral resolution at $\sim$7000 \AA, roughly double compared to
MaNGA's and one of the highest resolutions achieved by the surveys featured in Table \ref{tab:surveys}. Only the DiskMass survey
features a higher resolution with $R \sim 10,000$ at 6700 \AA.

\begin{deluxetable*}{lcccccl}
\tablecolumns{6}
\tablewidth{0pc}
\tablecaption{Prototype Run Plates \label{tab:fields}}
\tabletypesize{\footnotesize}
\tablehead{
\colhead{Field}  &\colhead{PlateID} &\colhead{R.A.} &\colhead{Decl.} & \colhead{$t_{\rm exp}$} & \colhead{Seeing} & \colhead{Comments} \\
\colhead{} &\colhead{} &\multicolumn{2}{c}{(J2000 degrees)} & \colhead{(hr)}  & \colhead{(\arcsec)}  & \colhead{}  \\
}
\startdata

\cutinhead{Galaxy Plates}
9 & 6650 & 143.74001& 21.788594   & 3.0  &  1.7 & Full depth; 3 sets of 3 dithers \\
11 & 6651 & 207.87862& 14.175544 & 1.0 & 2.0 & Shallow depth; 1 set of 3 dithers, airmass $\sim$1.5 (high) \\  
4 &  6652 & 163.98026& 36.944852  & 2.0 & 1.3 & Intermediate depth; 2 sets of 3 dithers \\

\cutinhead{All-sky Plates}
sky1 & 6613 & 161.011708 & 19.885278 & 1.7  & N/A & non-photometric, $\sim$full moon at $\Delta r \approx 40^\circ$ \\
sky2 & 6614 & 134.779583 & 3.832972 & 1.3  & N/A & photometric, full moon at $\Delta r \approx 8^\circ$ \\

\cutinhead{Star Plates}
Std-01 & 6653 & 188.657083 & 19.621389 & 1  & 1.6 & Flux standards \\
Std-01 & 6870 & 212.00000  &  18.000000 & 0.75  & 1.7 & Flux standards \\
PSF-01 & 6655 & 191.00000 & 18.000000 & 0.5  & 1.1 & Pointing, dithering, PSF characterization \\

\enddata
\tablecomments{RA and Dec values are reported in J2000 decimal degrees.}
\end{deluxetable*}

\section{P-M{\tiny a}NGA: Prototype Observations and Science Preview}\label{sec:results}

In this section we describe observations obtained with the P-MaNGA prototype instrument and present examples of early science
verification studies. Further results based on the P-MaNGA data will be presented in a series of three papers. \citet{belfiore14}
explore resolved maps of key emission line ratios to constrain the ionization state of gas in the P-MaNGA sample,
possible excitation mechanisms, and the inferred gas-phase abundance of key elements such as oxygen and nitrogen. D.~Wilkinson et
al.~(in preparation) develop and apply a new spectral fitter, {\sc firefly}, to the P-MaNGA datacubes, producing maps of $M_*$,
stellar age, metallicity, and dust. Finally, C.~Li et al.~(in preparation) take advantage of the high throughput at 4000 \AA\ to
study gradients in diagnostics of recent star formation activity, demonstrating the potential for future constraints on the nature
of star formation quenching (see Figure \ref{fig:quenching}).

\begin{figure*}
\epsscale{0.97}
\plotone{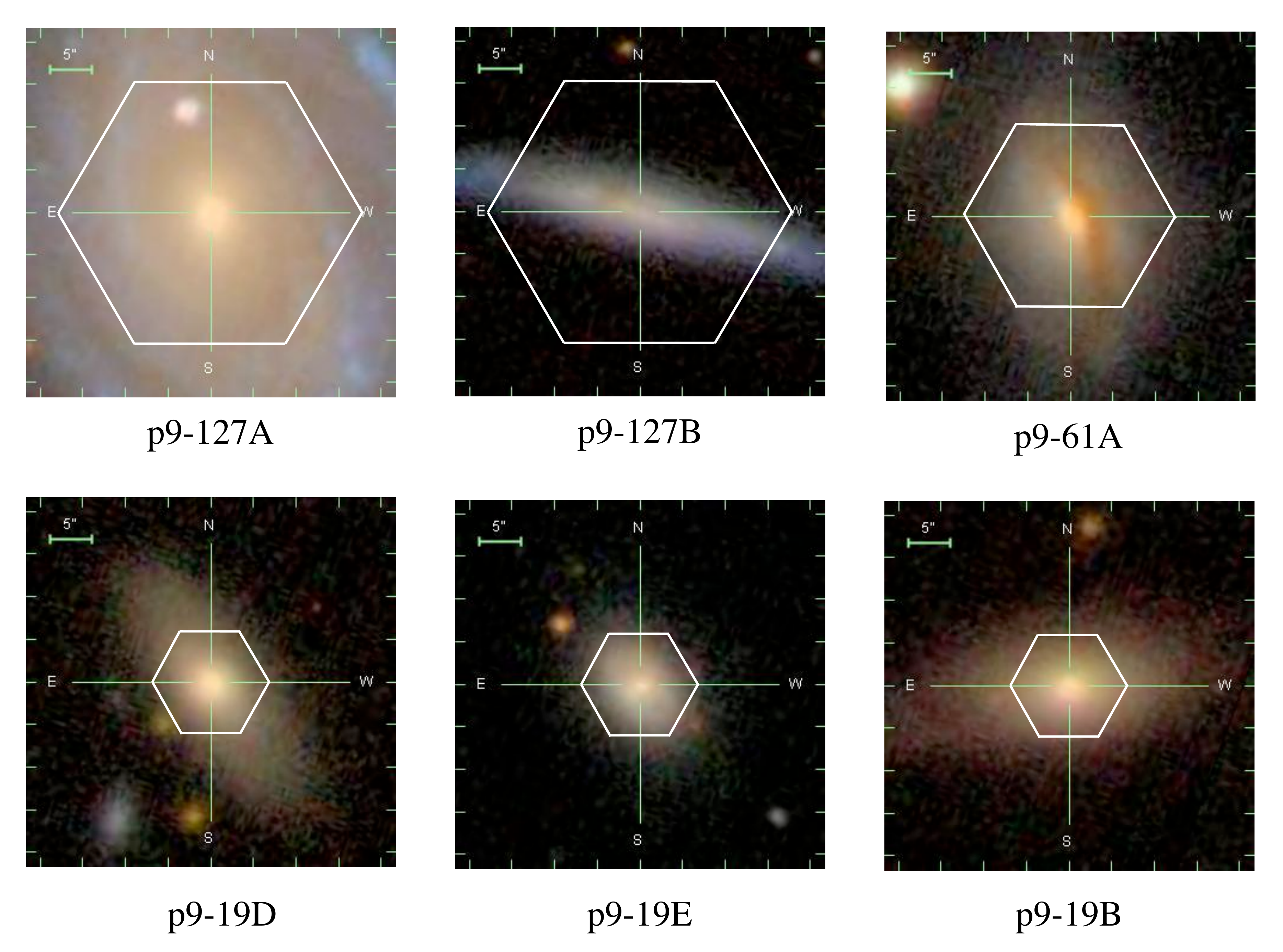}
\caption{SDSS $gri$ color image cutouts of the galaxies targeted in Field 9 and observed to full depth during the prototype run.  The hex pattern of the prototype IFU is overlaid, and the prototype run object name is given.  More details are given in Table \ref{tab:sample}. \label{fig:plate9_montage}}
\end{figure*}

\begin{deluxetable*}{lccccccccccl}
\tablecolumns{12}
\tablewidth{0pc}
\tablecaption{ Prototype Run Galaxies Observed \label{tab:sample}}
\tabletypesize{\footnotesize}
\tablehead{
\colhead{Name} &\colhead{mangaID} & \colhead{Bundle}  &\colhead{IAU Name} &\colhead{$z$} &\colhead{$\log$} &\colhead{$M_i$} &\colhead{$(g-r)$} &\colhead{$R_e$} &\colhead{Type$^{a}$} &  \colhead{$R_{\rm IFU}$}   &  \colhead{Comments} \\
\colhead{} & \colhead{} & \colhead{} &\colhead{} & \colhead{} & \colhead{$M_*$}& \colhead{(mag)} & \colhead{(mag)} & \colhead{('')} & \colhead{} & \colhead{(\Reff)} & \colhead{}   \\
}

\startdata

\cutinhead{Field 9: PlateID = 6650, 3.0 hr, seeing 1\farcs7}
p9-127A & 8-188794 &ma003      & J093457.30+214220.9 & 0.013 & 10.7 & -21.3 & 0.70 & 23.7 & $\star$  & 0.7   & NCG 2916; Face-on  \\
&  & & & & & & & & &  & spiral; CALIFA \\
p9-127B & 8-131835 & ma008 & J093506.31+213739.5 & 0.013 & 9.1 & -18.5 & 0.51 & 6.8 & 1 & 2.4  & CGCG122-022; Edge-on \\
p9-61A & 8-188807  & ma002   & J093712.30+214005.0 & 0.019 & 10.1 & -20.5 & 1.30 & 9.3 & $\star$ & 1.2   &  UGC 5124; E/S0, dust \\
p9-19B &8-131893  & ma005 & J093109.60+224447.4 & 0.051 & 10.6 & -22.2 & 0.86 & 4.0 & 1 & 1.6   & E/S0  \\
p9-19D &8-131577 & ma007 & J093109.07+205500.5 & 0.034 & 10.3 & -21.2 & 0.79 & 3.2 & 1 & 2.0   & E/S0\\
p9-19E &8-131821 & ma001 & J094030.22+211513.7 & 0.024 & 9.7 & -20.0 & 0.73 & 2.9 & 1 & 2.2  & Sa \\

\cutinhead{Field 11: PlateID = 6651, 1.0 hr, seeing 2\farcs0, (airmass $\sim$1.5)}
p11-127A &8-196354 & ma003   & J135133.10+140515.0 & 0.024 & 10.9 & -22.4 & 0.85 & 17.0 & $\star$ & 1.0    & IC 0944; Edge-on; \\
& & & & & & & & & &  &  CALIFA; dust \\
p11-127B &8-196727 & ma008 & J135655.69+140833.0 & 0.016 & 9.4 & -19.4 & 0.42 & 9.6 & $\star$ &1.7   & KUG 1354+143; Flocc. \\
p11-61A &8-93688 & ma002     & J135211.55+135959.6 & 0.024 & 10.1 & -20.7 & 0.77 & 7.3 & 1 & 1.5   & CGCG 073-090; red disk; \\
& & & & & & & & &  & & inclined; wants N$_{91}$\\
p11-19A &8-196773 & ma004& J135019.40+140822.9 & 0.024 & 9.5 & -21.1 & 0.85 & 1.1 & $\star$ & 2.8    & Edge-on disk \\
p11-19B &8-93551 & ma005 & J134910.01+132049.3 & 0.024 & 9.2 & -19.4 & 0.42 & 2.4 & 1 & 2.6    & Face-on blue spiral\\
p11-19C &8-93538 & ma006 & J134649.45+142401.7 & 0.021 & 10.0 & -20.9 & 0.54 & 2.8 & $\star$ & 2.2    & strong bar \\

\cutinhead{Field 4: PlateID = 6652, 2.0 hr, seeing 1\farcs3}
p4-127A &8-109661 & ma003 & J105555.26+365141.4 & 0.022 & 10.7 & -22.1 & 0.84 & 10.3 & $\star$  &1.6    & UGC 6036; Edge-on;  \\ 
& & & & & & & & &  & & CALIFA; dust \\
p4-127B &8-109682 & ma008    & J105259.05+373648.2 & 0.042 & 11.0 & -22.7 & 0.77 & 13.9 & $\star$ & 1.2  & Face-on spiral\\
p4-61A &8-113576 & ma002   & J105746.61+361657.8 & 0.030 & 9.7 & -20.0  & 0.71 & 3.4 & $\star$ & 3.4   & red core \\
p4-19A &8-113557 & ma004 & J110012.10+362313.8 & 0.027 & 9.4 & -19.5 & 0.49 & 2.5 &  1 & 2.5   & Blue spiral, bar?\\
p4-19B &8-113506 & ma005& J104958.69+362454.0 & 0.023 & 9.5 & -19.5 & 0.53 & 4.6 & 1 & 1.4    & Edge-on disk, wants N$_{37}$ \\
p4-19C &8-109657 & ma006& J105605.68+365736.1 & 0.022 & 9.5 & -19.4 & 0.72 & 4.8 &1 & 1.3    & E/S0, wants N$_{37}$ \\

\enddata
\tablecomments{The P-MaNGA Name is composed of a ``p'' for ``prototype'' followed by the field ID, a hyphen, and then the
  shorthand ID for the bundle used (see Table \ref{fig:ifu}).  This shorthand includes a
  number corresponding to N$_{\rm IFU}$ for that bundle.  The $(g-r)$ magnitude is extinction-corrected. \\
  $^{a}$ Target type 1 indicates the galaxy would be selected in the MaNGA Survey's Primary sample. A ``$\star$'' indicates a
  galaxy that was chosen by hand for the prototype run and does not pass the nominal selection criteria.}
\end{deluxetable*}

\begin{figure*}
\epsscale{0.95}
\plotone{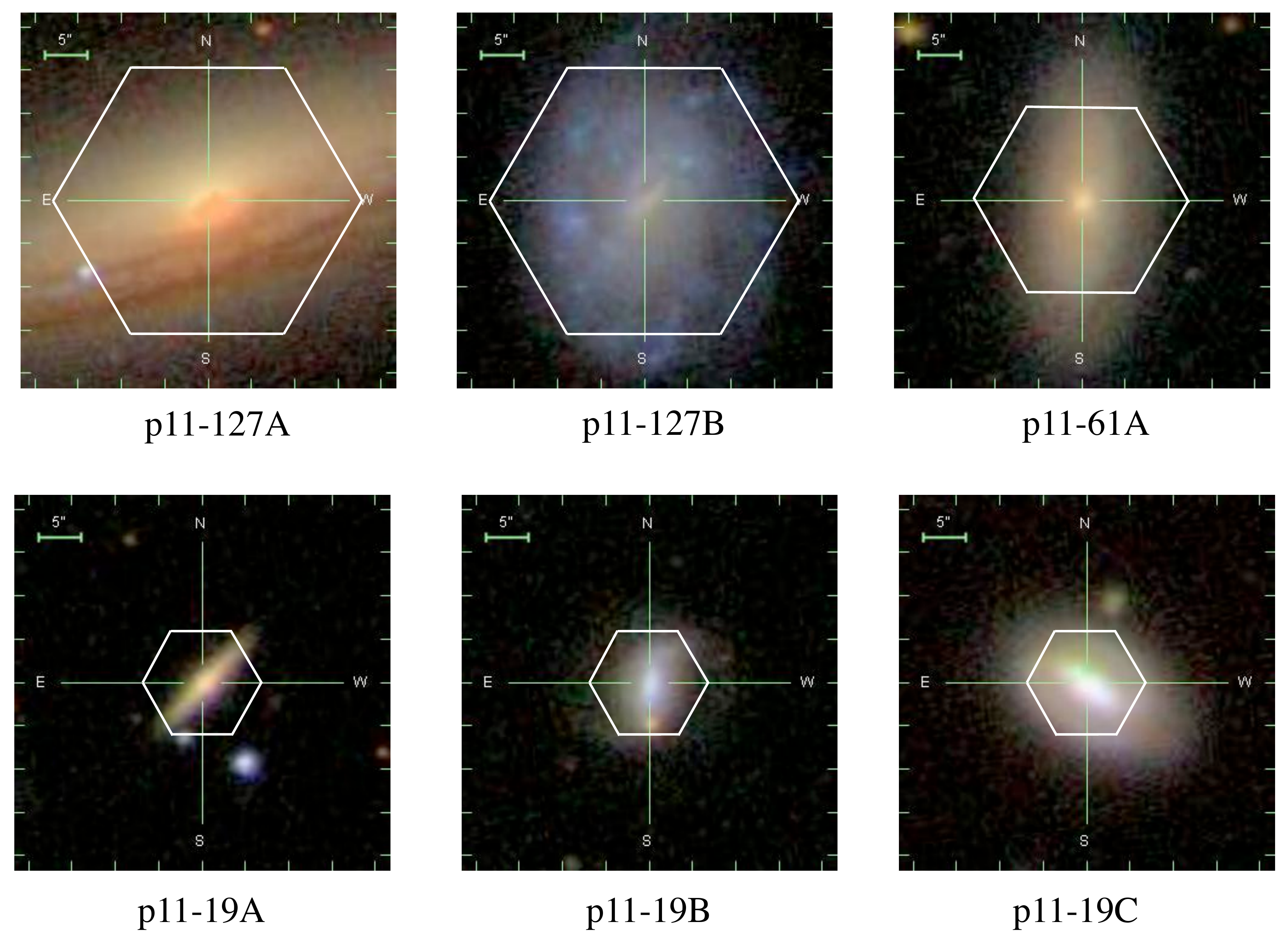}
\caption{SDSS $gri$ color image cutouts of the galaxies targeted in Field 11 during the prototype run.  The P-MaNGA observations of
  these galaxies reached $\sim$\sfrac{1}{3} the nominal depth and were conducted at relatively high airmass (1.5).  The hex pattern of the prototype IFU is overlaid, and the prototype run object name is given.  More details are given in Table \ref{tab:sample}. \label{fig:plate11_montage}}
\end{figure*}

\begin{figure*}
\epsscale{0.95}
\plotone{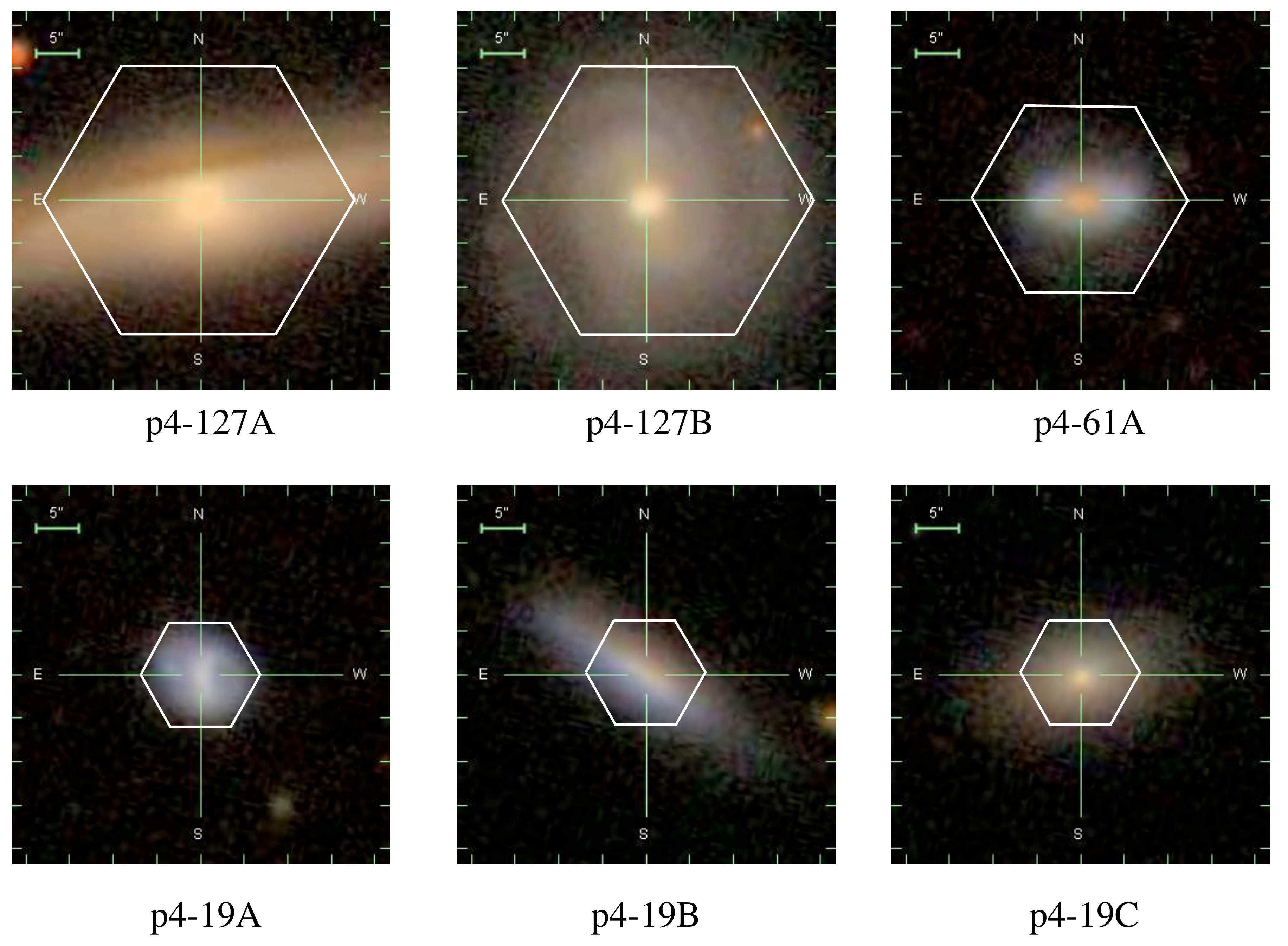}
\caption{SDSS $gri$ color image cutouts of the galaxies targeted in Field 4 and observed to $\sim$\sfrac{2}{3} the nominal depth during the prototype run.  The hex pattern of the prototype IFU is overlaid, and the prototype run object name is given.  More details are given in Table \ref{tab:sample}. \label{fig:plate4_montage}}
\end{figure*}

\subsection{Description of Observations}\label{sec:obs}

P-MaNGA observations in 2012 December and 2013 January were made possible by a donation of $\sim$11 clear hours on the Sloan
Telescope by the SDSS-III collaboration.  In addition to daytime calibration data (arcs and flats), we conducted a number of
experiments and observations using eight custom-drilled plug-plates (see Table \ref{tab:fields}).  We began with ``all-sky'' plates
(6613 and 6614) on which all fibers and IFUs were located in empty sky regions.  These plates were observed during bright time and,
in the case of 6614, close to a nearly full moon.  These observations were used to test the performance of sky subtraction
algorithms by using subsets of the data to build a sky model, which was then subtracted from the remaining IFU fibers.  Studies of
the residual signal (which under perfect subtraction would have been zero everywhere) provided a valuable means for quantifying
sky subtraction performance.  These studies led to the discovery of the coma aberration in the red channel of the SP1 BOSS
spectrograph and to improvements in MaNGA's final design (see \S\ref{sec:p-manga-hardware}).

A second set of experiments was performed by placing fibers and IFUs predominantly on stars.  Two ``standard-star'' plates were
observed to test various options for achieving MaNGA's requirements on spectrophotometric calibration.  We also observed a special
``PSF plate'' on which several IFUs were positioned in order to simultaneously capture close groupings of two to three stars within the
IFU's field of view.  By dithering observations with this plate, we could pinpoint the position and orientation of the IFUs, as
well as the precision to which offset dithers could be made.  We were also able to test our ability to recover the atmospheric
PSF from the fiber-sampled IFU data---an important ingredient in the flux calibration---as well as the importance of
fiber-to-fiber cross talk on the detector for the precision of the inferred PSF.

Finally, we observed three galaxy fields (referred to for the prototype run as 9, 11, and 14) using plates 6650, 6651, and 6652.
In each case, observations were obtained in sets of three 20-minute exposures dithered by roughly a fiber radius along the
vertices of an equilateral triangle to provide uniform coverage across each IFU.  These three fields were observed to varying
depths and in varying conditions as required by the P-MaNGA engineering tasks for which they were designed.  Although plate 6650
(Field 9) was observed to a depth comparable to what will be regularly achieved during MaNGA operations, plates 6651 (Field 11)
and 6652 (Field 4) are both significantly shallower than MaNGA survey data, and plate 6651 was intentionally observed at high
airmass, resulting in particularly poor image quality.

The MaNGA Survey sample design was reviewed in \S\ref{sec:survey}.  Some of the P-MaNGA targets were drawn from early versions of
the MaNGA selection, but in many cases P-MaNGA targets were chosen for specific reasons.  SDSS $gri$ images of the targets from
each P-MaNGA field are shown in Figures \ref{fig:plate9_montage} through \ref{fig:plate4_montage}. In each of these fields, one
\N{127} IFU was allocated to a galaxy observed by the CALIFA survey \citep{sanchez12},
even if it would not otherwise satisfy the MaNGA selection cuts.  Additionally, the nonoptimal IFU complement of the P-MaNGA
instrument required some targets to be selected by hand, and, on each plate, one \N{19} IFU was assigned to a blank-sky region and
one to a standard star.  Altogether 18 galaxies were observed (see Table \ref{tab:sample}) in P-MaNGA.  Of these, only four
(p9-127B, p9-19B, p9-19D, and p9-19E) have the exposure time, spatial resolution, and radial coverage expected for MaNGA survey
observations for IFUs with the associated size.

\begin{figure*}[ht!]
\includegraphics[width=1\textwidth]{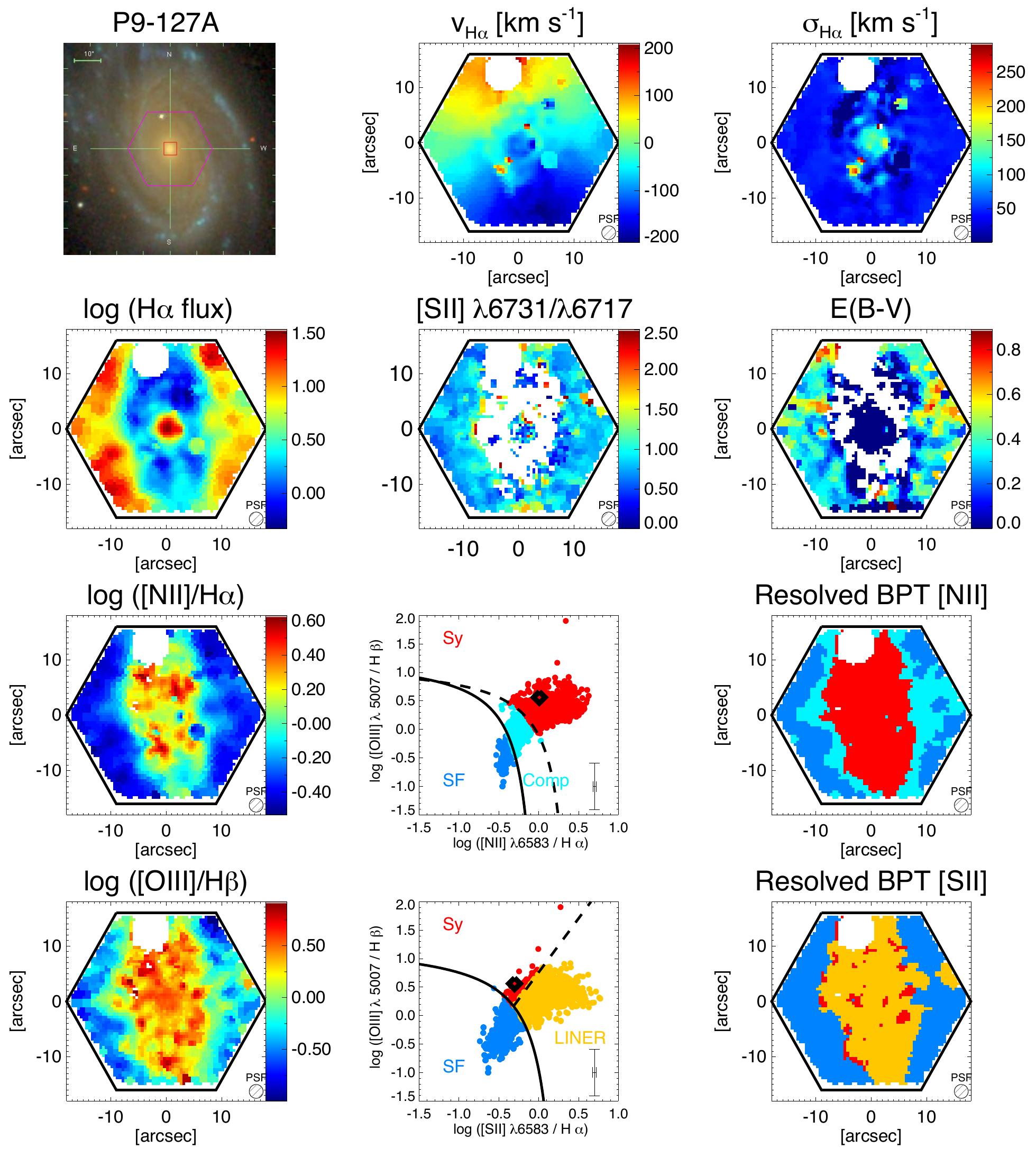} 
\caption{Panoramic view of resolved, ionization diagnostic information available for MaNGA targets, taken from \citet{belfiore14}.  Results for the P-MaNGA target, P9-127A (NGC 2916), are shown.  The black diamonds in the BPT diagrams
  represent the value obtained in the central fiber.  Flux measurements are given in units of 10$^{-17}$ erg s$^{-1}$ cm$^{-2}$
  \AA$^{-1}$.  Central white spaces are
  regions where one or more emission lines are too weak to provide reliable ratios or diagnostic measurements.   A foreground star in
  the upper portion of the IFU data has been masked.  See text
  (\S\ref{sec:ionization}) for details.}
\label{fig:ion}
\end{figure*}

The raw data were processed using a prototype of the MaNGA Data Reduction Pipeline (DRP), which is described in detail by D.~R.~Law et
al.~(in preperation).  In brief, individual fiber flux and inverse variance spectra were extracted using a row-by-row algorithm,
wavelength calibrated using a series of neon-mercury-cadmium arc lines, and flat-fielded using internal quartz calibration lamps.
Sky subtraction of the IFU fiber spectra was performed by constructing a cubic basis spline model of the sky background flux as
seen by the 41 individual fibers placed on blank regions of sky (the remaining 19 2\arcsec\ fibers were assigned to standard
stars) and subtracting the resulting composite spectrum shifted to the native wavelength solution of each IFU fiber.  Primarily
limited by the SP1 coma, bright skyline residuals at the reddest wavelengths remain at the 3--4\% level.  At a reference
wavelength of 9378 \AA, this residual corresponds to roughly 2--3 times the theoretical limit as set by the Poisson error expected
from the observed skyline flux.  The Poisson limit was achieved, however, blueward of 7000~\AA.

\begin{figure*}[ht]
  \epsscale{1.0}
  \plottwo{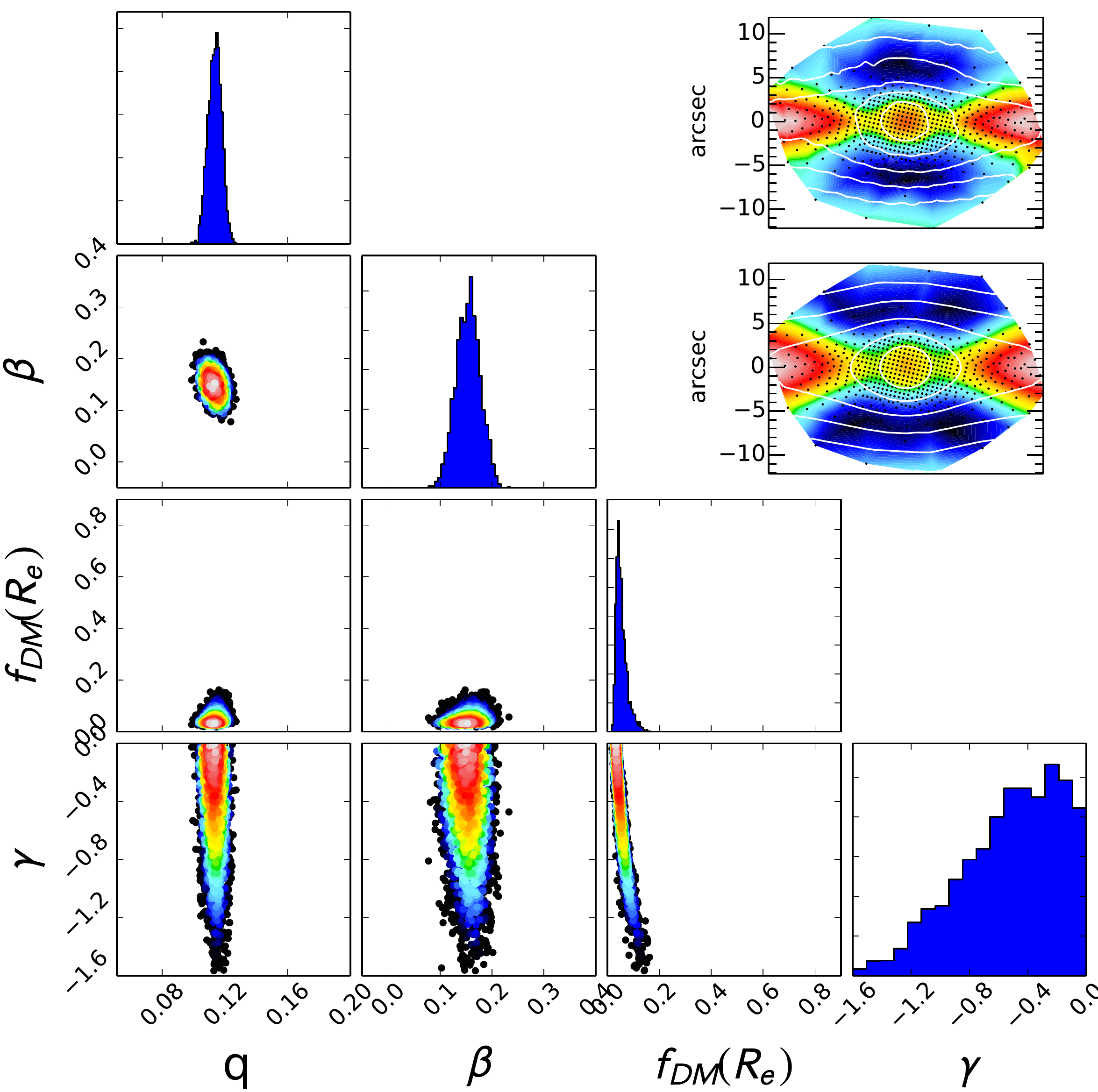}{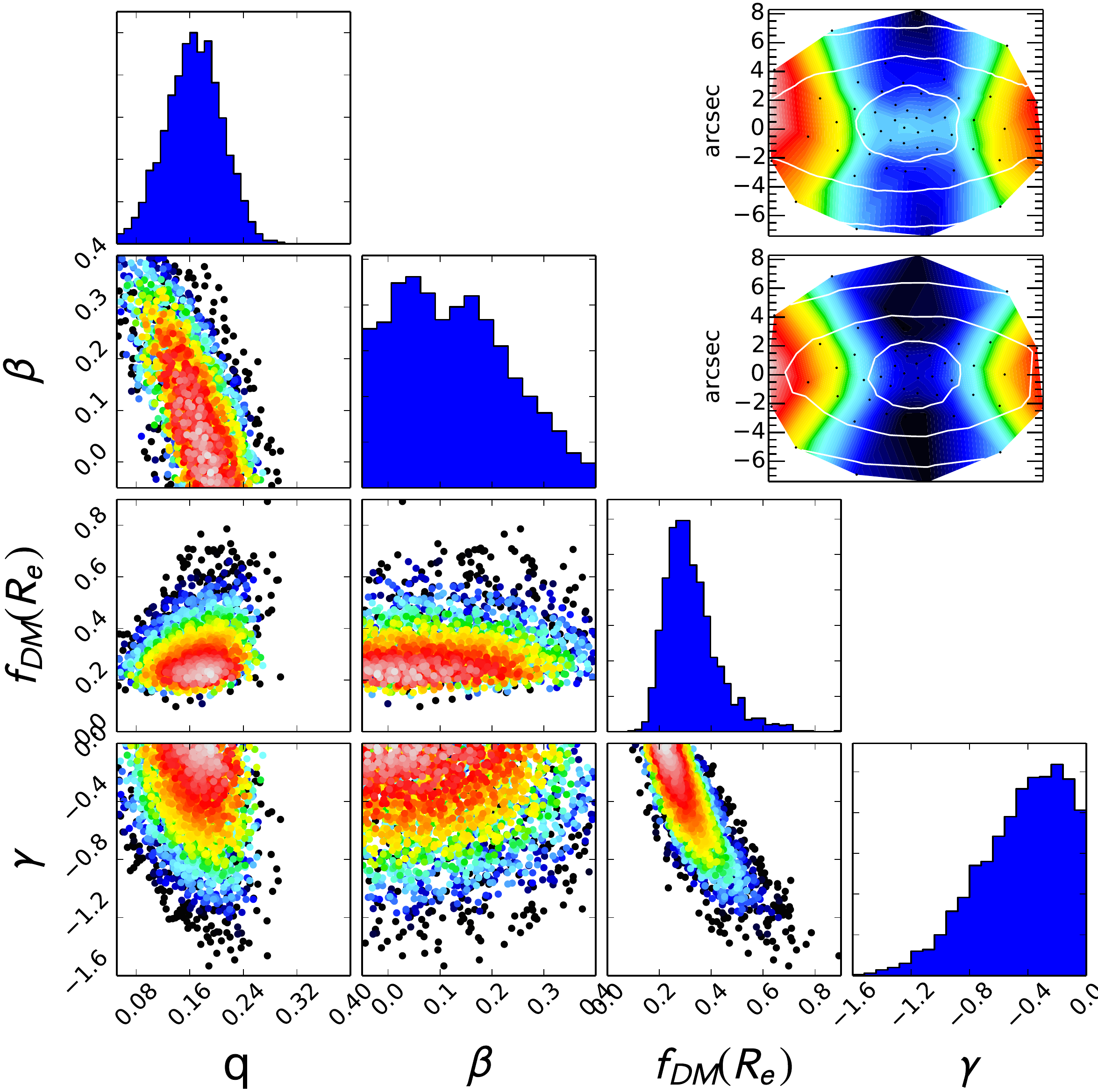} 
  \caption{Examples of JAM dynamical modeling of two MaNGA prototype galaxies. The left galaxy (p4-127A) is one of the best fits in
    the prototype sample, while the one on the right, p11-61A, is of poor quality and not representative of MaNGA survey
    observations (it was observed for only 1 hr at high airmass and poor seeing).  Each panel displays the posterior probability
    distribution for the nonlinear model parameters: (a) the intrinsic axial ratio $q$; (b) the anisotropy $\beta_z$; (c) the
    dark matter fraction $f_{\rm DM}(R_e)$ within a sphere of radius $R_e$; and (d) the inner dark halo logarithmic slope
    $\gamma$. The posterior was marginalized both over two dimensions (color contours) and in one dimension (blue histograms). The
    symbols are colored according to their likelihood: white corresponds to the maximum value and dark blue to a $3\sigma$
    confidence level. For each combination of the nonlinear parameters, the linear parameter $(M/L)_{\rm stars}$ is adjusted to
    fit the data.  In insets above and to the right of the probability distributions, the symmetrized $V_{\rm
      rms}\equiv\sqrt{V^2+\sigma^2}$ and linearly interpolated MaNGA data are shown, with the best-fitting model reproduced
    below. Observed and model surface brightness contours are overlaid, in 1.0 mag intervals. The centroid of the Voronoi bins is
    shown with the small dots. The lower projected density of Voronoi cells in p11-61A is an indication of its poorer data quality.}%
\label{fig:jam}%
\end{figure*}

Flux calibration of the P-MaNGA data is performed by fitting Kurucz model stellar spectra to the spectra of calibration standard
stars covered with single fibers at each of the three dither positions.  The flux calibration vectors derived from these
single-fiber spectra were found to vary by $\sim$ 10\% from exposure to exposure, depending on the amount of light lost from the
fiber due to atmospheric seeing and astrometric misalignments.  While this uncertainty is acceptable for the present science
purposes, the flux calibration uncertainty of the single fibers ultimately drove the design decision of the MaNGA survey to
use seven-fiber IFU ``mini-bundles'' for each calibration standard star, which results in a 2--3\% photometric uncertainty
(see R.~Yan et al., in preperation).

Flux-calibrated spectra from the blue and red cameras were combined across the dichroic break using an inverse-variance weighted
basis spline function. Astrometric solutions were derived for each individual fiber spectrum that incorporate information about
the IFU bundle metrology (i.e., fiber location within an IFU), dithering, and atmospheric chromatic differential refraction, among
other effects. Fiber spectra from all exposures for a given galaxy were then combined into a single datacube (and corresponding
inverse variance array) using these astrometric solutions and a nearest-neighbor sampling algorithm similar to that used by the
CALIFA survey. For the P-MaNGA datacubes, a spaxel size of 0\farcs5 was chosen. The typical effective spatial resolution in the
reconstructed datacubes can be described by a Gaussian with ${\rm FWHM} \approx 2$\farcs5. When binning the datacubes, the
resulting error vectors are scaled to approximately account for wavelength and spatial covariance in the P-MaNGA error cubes. A
more accurate estimation of spectral error vectors and their propagation through to reported errors for each datacube will be
developed in the first year of the survey.

\subsection{Early Science Analyses}\label{sec:ionization}\label{sec:ion}\label{sec:dyn}

We present several preliminary analyses that demonstrate the utility of the P-MaNGA observations and the scientific potential of
MaNGA survey data.

In Figure \ref{fig:ion} we show line ratio maps that probe the ionization state of P-MaNGA galaxy P9-127A (NGC 2916).  Resolved
maps of the classical Baldwin-Phillips-Terlevich (BPT; \citealt{baldwin81, veilleux87}) diagram composed of the line ratio
\OIII/H$\beta$ versus \NII/H$\alpha$ and \SII/H$\alpha$, are a valuable way of distinguishing ionization sources
\citep[e.g.,][]{sharp10}. The \NII\ BPT diagram is color-coded by the inferred ionization source \citep[e.g.,][]{kewley06} and the
same color-coding is adopted to map the position of these regions within the galaxy, which we show together with the maps of the
line ratios \NII/H$\alpha$ and \OIII/H$\beta$. The black hexagon corresponds to the MaNGA bundle field of view.  In this galaxy,
the central regions are photoionized by a Seyfert-like radiation field according to the rough demarcation from \citet{kewley06},
suggesting a high ionization parameter ($\log u$) coupled with a drop in the $\mathrm{ H \alpha}$ flux everywhere except in the central
bulge.  Overall the galaxy features a massive disk, with extensive star formation in the outskirts that was missed in the Sloan
$3''$ fiber (represented in the BPT diagram as an open diamond).  Further details and a full analysis of the P-MaNGA sample are
presented in \citet{belfiore14}.

We also demonstrate how the P-MaNGA data provide physical insight from dynamical information. Figure \ref{fig:jam} shows
successful Jeans axisymmetric modelling \citep[JAM,][]{cappellari2008} applied to two P-MaNGA galaxies, with derived constraints
on the intrinsic axis ratio, anisotropy, dark matter fraction (assuming a generalized Navarro-Frenk-White (NFW), with free inner slope), and inner dark
matter slope. Tests performed on the full sample indicate that the separation of stellar and dark components will be possible not
only for a small subset of MaNGA survey galaxies, where {\em random} errors in the stellar mass-to-light ratio will be on
the order of a few percent, but for the majority of MaNGA galaxies, where {\em random} errors on the order of 10\% are expected,
given our assumptions about the dark matter halo. Assessing the impact of the likely larger systematic errors that result from
these and other assumptions is a major subject of investigation for future work.

\begin{figure}
\epsscale{1.15}
\plotone{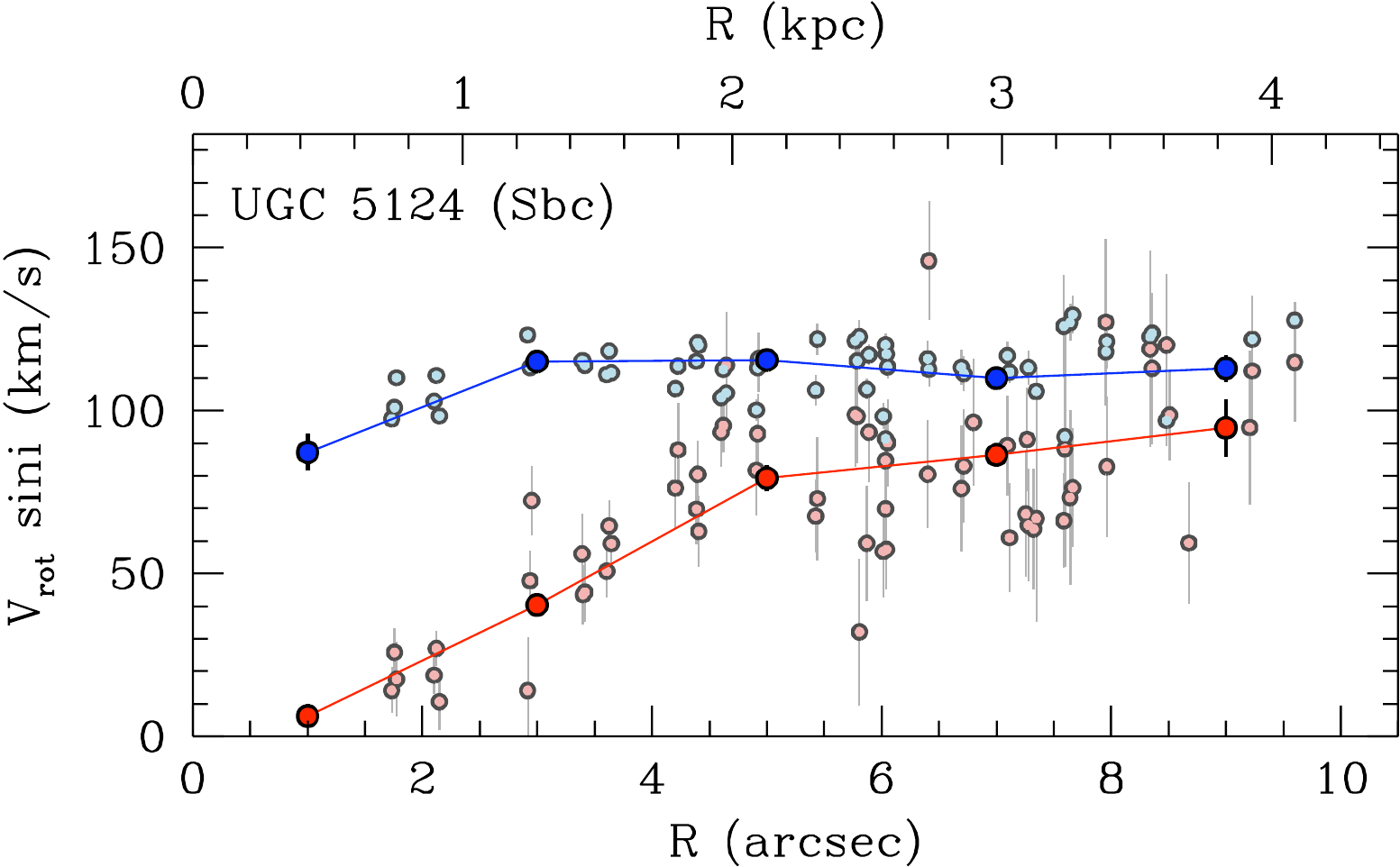}

\caption{Detection of asymmetric drift in the MaNGA prototype observations of UGC 5124 (p9-61A), which has an inclination of
    $i=45\fdg6\pm2\fdg0$ and a distance of 87.8 Mpc. Small, light-blue and light-red points are measurements of ionized-gas and stellar velocities,
    respectively, as a function of disk-plane radius ($R$) from individual fibers that have been projected onto the major axis of
    rotation.  Only those data within $\pm60\arcdeg$ (disk-plane azimuth) of the major axis have been plotted.  Large blue and
    red points show the error-weighted mean, projected rotation speed ($V_{\rm rot} \sin i$) in $2\arcsec$ bins for the
    ionized-gas and stellar components, respectively.  As expected, the stellar (red) component lags behind the ionized-gas (blue)
    component owing to the substantially larger velocity dispersion and collisionless nature of the stars.
\label{fig:AD}}
\end{figure}


We have also detected asymmetric drift in the dynamics of P-MaNGA galaxies.  An example is shown in Figure \ref{fig:AD}.
This phenomenon manifests as a lag in the rotation speed of stars compared to the gas because the stars, which are collisionless,
maintain a velocity distribution function that has nonzero width, i.e., some of the kinetic energy of the stellar system is tied
up in a pressure term. Gaseous tracers, on the other hand, are collisional and can dissipate energy and settle into a dynamically
cold disk with a rotation speed that approximately follows the circular speed \citep[e.g.,][]{dalcanton10}.  Given the direct
relation between AD and the velocity distribution function, AD can be used to infer the dynamical mass surface density
\citep[e.g.,][K.~B.~Westfall et al.~in preparation]{van-der-kruit88, bershady10}, the disk stability level \citep[e.g.,][]{romeo13,
  westfall14}, and highly accurate measurements of inclination, especially for nearly face-on systems \citep{blanc13a}.

\begin{figure}[ht!] 
\epsscale{1.25}
\plotone{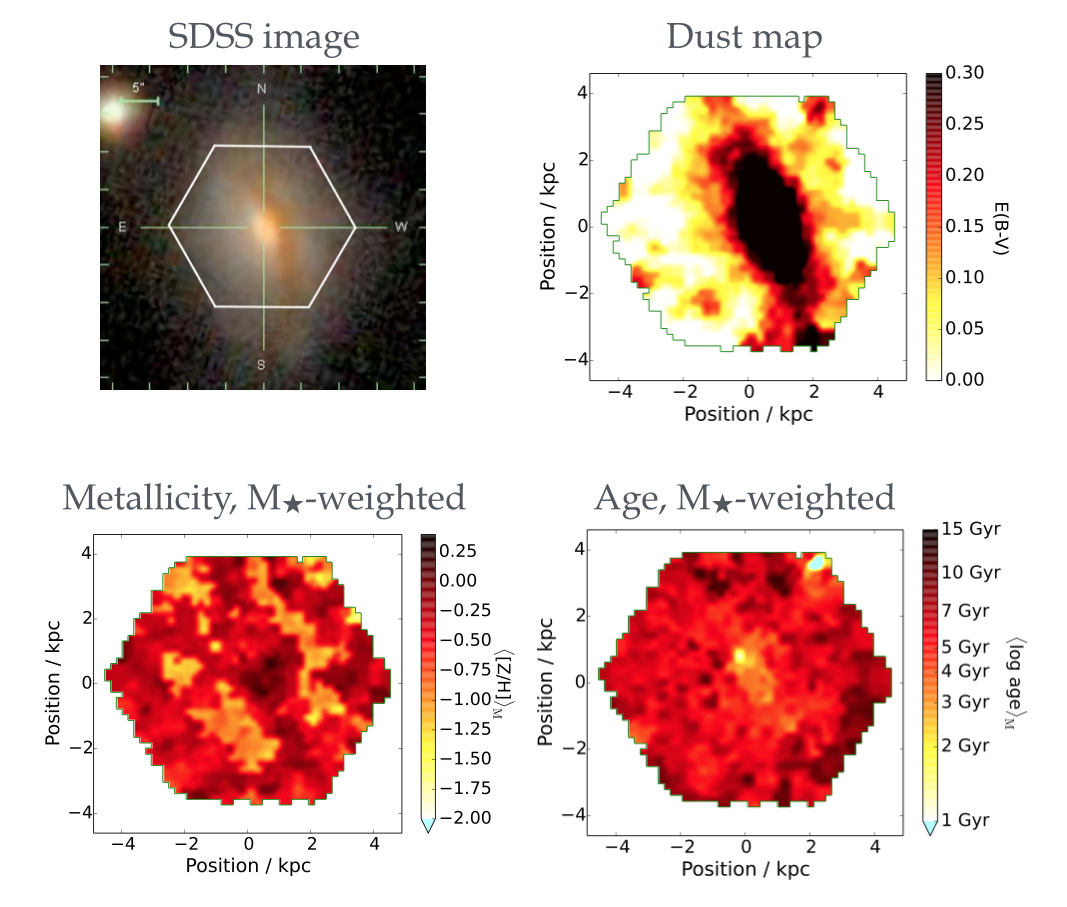}
\caption{Stellar population maps of P-MaNGA galaxy p9-61A, produced with the {\sc firefly} spectral fitting code (D.~Wilkinson, in preparation)
  using MILES-based models. The mass-weighted metallicity and age maps (lower panels) are free of dust contamination, which is
  measured in the upper-right panel. \label{fig:wilkinson1}}
\end{figure}

Finally, in Figure \ref{fig:wilkinson1} we show maps of stellar population parameters derived from the {\sc firefly} spectral
fitter for P-MaNGA galaxy P9-61A. This example highlights how spectral fitting applied to MaNGA data can measure dust extinction
and break the dust--age--metallicity degeneracy. The derived dust map (upper right) is aligned with the dust lane apparent in the
image, but the mass-weighted stellar metallicity and stellar age maps (lower panels) show little coincident structure in this
region (offset just northwest of the nucleus), while indicating a younger and somewhat more metal-rich central population, which may be
related to more recent star formation associated with the dust lane. The ability to break such degeneracies would be extremely
challenging without resolved spectroscopy. Further details are provided in D.~Wilkinson et al.~(in preperation).

\section{Conclusions}\label{sec:conclusions}

We have presented an overview of the MaNGA instrument and survey concept, discussed its motivation and design with respect to past
and ongoing IFU surveys, and used observations obtained with a prototype of the final instrument to demonstrate MaNGA's scientific
potential. The MaNGA survey is part of the SDSS-IV project which officially began on 2014 July 1. Over the coming 6 yr,
MaNGA will obtain resolved spectroscopy for an unprecedented sample of 10,000 galaxies.  

MaNGA employs 17 science IFUs over the 2.5m Sloan Telescope's 3$^{\circ}$ diameter field of view.  These IFUs vary in size from
12\arcsec\ to 32\arcsec\ in diameter, with a distribution that is matched to the apparent size of galaxy targets on the sky.
Twelve mini-bundles are used to target standard stars and provide relative flux calibration at a precision of a few percent
between 3700 \AA\ and 6500 \AA.  We employ 92 sky fibers to achieve nearly Poisson-limited sky subtraction.  The MaNGA fibers feed
the two BOSS spectrographs, each with a red and blue channel that provide simultaneous wavelength coverage from 3600 to 10,300 \AA\
with a mid-range resolution of $R \sim 2000$.  

MaNGA's main samples are selected from SDSS-I using only redshifts and $i$-band luminosities in order to achieve nearly uniform
radial coverage to radii of 1.5 \Reff\ for two-thirds of the final sample (the Primary sample) and 2.5 \Reff\ for one-third
of the final sample (the Secondary sample).  This is accomplished with a luminosity-dependent, volume-limited selection.  An
additional ``color enhancement,'' equal to 25\% of the Primary sample, provides nearly flat distributions in both $M_*$ and color
for the combined, Primary+ sample.

MaNGA employs dithered exposures and observes until target S/N values are reached for each plate.  These thresholds correspond to
$r$-band S/N values of 4--8 \AA\per\ per fiber in the outskirts of Primary sample galaxies and typically require 3--4 hr of
observations under median conditions.

As with past SDSS surveys, MaNGA data and data products will be made public through ``Data Releases'' (DRs) occurring approximately every
year throughout the course of the project. The initial released MaNGA data will
include raw spectroscopic frames and reduced, flux-calibrated, sky-subtracted fiber spectra with associated position
look-up tables specific to the corresponding IFU and set of dithered observations for each targeted galaxy. Interpolated final
datacubes will also be included in early releases. In the future, more advanced data products including maps of spectral line
flux measurements and continuum diagnostics, as well as model-derived estimates of physical properties (e.g., SFR, metallicity,
kinematic maps), will also be made available.  One of MaNGA's goals is to publicly provide these data products using a
well-documented, interactive, and user-friendly platform that will enable MaNGA's complex data set to be fully exploited.

\section{Acknowledgments}

We are grateful for careful comments on the manuscript by Don Schneider and input from the SAMI, CALIFA, and \atl\ teams.  We also
thank Adam Bolton, Stephane Courteau, Natascha Forster-Schreiber, Bruce Gillespie, Gary Hill, Arlette P\'econtal, Rob Sharp, Ray
Sharples, and Dennis Zaritsky for valuable feedback on the MaNGA instrumentation and survey design.  Funding for SDSS-III and
SDSS-IV has been provided by the Alfred P.~Sloan Foundation and Participating Institutions.  Additional funding for SDSS-III comes
from the National Science Foundation and the U.S.~Department of Energy Office of Science.  Further information about both projects
is available at {\tt www.sdss.org}.

SDSS is managed by the Astrophysical Research Consortium for the Participating Institutions in both collaborations.  In SDSS-III
these include the University of Arizona, the Brazilian Participation Group, Brookhaven National Laboratory, Carnegie Mellon
University, University of Florida, the French Participation Group, the German Participation Group, Harvard University, the
Instituto de Astrofisica de Canarias, the Michigan State/Notre Dame/JINA Participation Group, Johns Hopkins University, Lawrence
Berkeley National Laboratory, Max Planck Institute for Astrophysics, Max Planck Institute for Extraterrestrial Physics, New Mexico
State University, New York University, Ohio State University, Pennsylvania State University, University of Portsmouth, Princeton
University, the Spanish Participation Group, University of Tokyo, University of Utah, Vanderbilt University, University of
Virginia, University of Washington, and Yale University.

The Participating Institutions in SDSS-IV are Carnegie Mellon University, Colorado University, Boulder, Harvard-Smithsonian Center
for Astrophysics Participation Group, Johns Hopkins University, Kavli Institute for the Physics and Mathematics of the Universe,
Max-Planck-Institut fuer Astrophysik (MPA Garching), Max-Planck-Institut fuer Extraterrestrische Physik (MPE), Max-Planck-Institut
fuer Astronomie (MPIA Heidelberg), National Astronomical Observatories of China, New Mexico State University, New York University,
The Ohio State University, Penn State University, Shanghai Astronomical Observatory, United Kingdom Participation Group,
University of Portsmouth, University of Utah, University of Wisconsin, and Yale University.

This work was supported in part by a World Premier International Research Center Initiative (WPI Initiative), MEXT, Japan.  J.H.K.\
acknowledges financial support to the DAGAL network from the People Programme (Marie Curie Actions) of the European Union's
Seventh Framework Programme FP7/2007-2013/ under REA grant agreement no.\ PITN-GA-2011-289313.  This work has also been
supported by the Strategic Priority Research Program ``The Emergence of Cosmological Structures'' of the Chinese Academy of
Sciences Grant No. XDB09000000 (S.M., C.L.), and by the National Natural Science Foundation of China (NSFC) under grant numbers 11333003
(S.M.).  M.C.\ acknowledges support from a Royal Society University Research Fellowship.  CC acknowledges support from the Sloan and
Packard Foundations.  A.M.D.\ acknowledges support from The Grainger
Foundation.

\bibliographystyle{apj}
\bibliography{manga_references}

\end{document}

%% file: authors.tex

\author{Kevin Bundy\altaffilmark{1}, Matthew A.~Bershady\altaffilmark{2}, David R.~Law\altaffilmark{3}, Renbin Yan\altaffilmark{4}, Niv Drory\altaffilmark{5}, Nicholas MacDonald\altaffilmark{6}, David A.~Wake\altaffilmark{2}, Brian Cherinka\altaffilmark{3}, Jos\'e R.~S\'anchez-Gallego\altaffilmark{4}, Anne-Marie Weijmans\altaffilmark{7}, Daniel Thomas\altaffilmark{8}, Christy Tremonti\altaffilmark{2}, Karen Masters\altaffilmark{8,9}, Lodovico Coccato\altaffilmark{10,8}, Aleksandar M.~Diamond-Stanic\altaffilmark{2}, Alfonso Arag\'on-Salamanca\altaffilmark{11}, Vladimir Avila-Reese\altaffilmark{12}, Carles Badenes\altaffilmark{13}, J\'esus Falc\'on-Barroso\altaffilmark{14}, Francesco Belfiore\altaffilmark{15,16}, Dmitry Bizyaev\altaffilmark{17}, Guillermo A.~Blanc\altaffilmark{18}, Joss Bland-Hawthorn\altaffilmark{19}, Michael R.~Blanton\altaffilmark{20}, Joel R.~Brownstein\altaffilmark{21}, Nell Byler\altaffilmark{6}, Michele Cappellari\altaffilmark{22}, Charlie Conroy\altaffilmark{23}, Aaron A.~Dutton\altaffilmark{24}, Eric Emsellem\altaffilmark{10,25}, James Etherington\altaffilmark{8}, Peter M.~Frinchaboy\altaffilmark{26}, Hai Fu\altaffilmark{27}, James E. Gunn\altaffilmark{28}, Paul Harding\altaffilmark{29}, Evelyn J.~Johnston\altaffilmark{11}, Guinevere Kauffmann\altaffilmark{30}, Karen Kinemuchi\altaffilmark{17}, Mark A.~Klaene\altaffilmark{17}, Johan H. Knapen\altaffilmark{14,31}, Alexie Leauthaud\altaffilmark{1}, Cheng Li\altaffilmark{32}, Lihwai Lin\altaffilmark{33}, Roberto Maiolino\altaffilmark{15}, Viktor Malanushenko\altaffilmark{17}, Elena Malanushenko\altaffilmark{17}, Shude Mao\altaffilmark{34,35}, Claudia Maraston\altaffilmark{8}, Richard M.~McDermid\altaffilmark{36,37}, Michael R.~Merrifield\altaffilmark{11}, Robert C.~Nichol\altaffilmark{8}, Daniel Oravetz\altaffilmark{17}, Kaike Pan\altaffilmark{17}, John K.~Parejko\altaffilmark{38}, Sebastian F.~Sanchez\altaffilmark{12}, David Schlegel\altaffilmark{39}, Audrey Simmons\altaffilmark{17}, Oliver Steele\altaffilmark{8}, Matthias Steinmetz\altaffilmark{40}, Karun Thanjavur\altaffilmark{41}, Benjamin A.~Thompson\altaffilmark{26}, Jeremy L.~Tinker\altaffilmark{20}, Remco C.~E.~van den Bosch\altaffilmark{24}, Kyle B.~Westfall\altaffilmark{42}, David Wilkinson\altaffilmark{8}, Shelley Wright\altaffilmark{3,43}, Ting Xiao\altaffilmark{32}, Kai Zhang\altaffilmark{4}}

\altaffiltext{1}{Kavli Institute for the Physics and Mathematics of the Universe (Kavli IPMU, WPI), Todai Institutes for Advanced Study, the University of Tokyo, Kashiwa 277-8583, Japan}
\altaffiltext{2}{Department of Astronomy, University of Wisconsin-Madison, 475 N. Charter Street, Madison, WI, 53706, USA}
\altaffiltext{3}{Dunlap Institute for Astronomy and Astrophysics, University of Toronto, 50 St. George Street, Toronto, Ontario M5S 3H4, Canada}
\altaffiltext{4}{Department of Physics and Astronomy, University of Kentucky, 505 Rose Street, Lexington, KY 40506-0055, USA}
\altaffiltext{5}{McDonald Observatory, Department of Astronomy, University of Texas at Austin, 1 University Station, Austin, TX 78712-0259, USA}
\altaffiltext{6}{Department of Astronomy, Box 351580, University of Washington, Seattle, WA 98195, USA}
\altaffiltext{7}{School of Physics and Astronomy, University of St Andrews, North Haugh, St Andrews KY16 9SS, UK}
\altaffiltext{8}{Institute of Cosmology and Gravitation, University of Portsmouth, Portsmouth, UK}
\altaffiltext{9}{SEPNet, South East Physics Network (www.sepnet.ac.uk), Southampton SO17 1BJ, UK}
\altaffiltext{10}{European Southern Observatory, Karl-Schwarzschild-Str. 2, D-85748 Garching, Germany}
\altaffiltext{11}{School of Physics and Astronomy, University of Nottingham, University Park, Nottingham NG7 2RD, UK}
\altaffiltext{12}{Instituto de Astronomia, Universidad Nacional Autonoma de Mexico, A.P. 70-264, 04510 Mexico D.F., Mexico}
\altaffiltext{13}{Department of Physics and Astronomy and Pittsburgh Particle Physics, Astrophysics and Cosmology Center (PITT PACC), University of Pittsburgh, 3941 O’Hara St, Pittsburgh, PA 15260, USA}
\altaffiltext{14}{Instituto de Astrof\'isica de Canarias, E-38200 La Laguna, Tenerife, Spain}
\altaffiltext{15}{Cavendish Laboratory, University of Cambridge, 19 J.~J.~Thomson Ave., Cambridge CB3 0HE, UK}
\altaffiltext{16}{Kavli Institute for Cosmology, University of Cambridge, Madingley Road, Cambridge CB3 0HA, UK}
\altaffiltext{17}{Apache Point Observatory and New Mexico State, P.O. Box 59, Sunspot, NM 88349, USA}
\altaffiltext{18}{Observatories of the Carnegie Institution for Science, Pasadena, CA, USA}
\altaffiltext{19}{Sydney Institute for Astronomy (SIfA), School of Physics, University of Sydney, NSW 2006, Australia}
\altaffiltext{20}{Center for Cosmology and Particle Physics, Department of Physics, New York University, 4 Washington Place, New York, NY 10003, USA}
\altaffiltext{21}{Department of Physics and Astronomy, University of Utah, 115 S 1400 E, Salt Lake City, UT 84112, USA}
\altaffiltext{22}{Sub-Department of Astrophysics, Department of Physics, University of Oxford, Denys Wilkinson Building, Keble Road, Oxford OX1 3RH, UK}
\altaffiltext{23}{Department of Astronomy and Astrophysics, University of California, Santa Cruz, CA 95064, USA}
\altaffiltext{24}{Max Planck Institute for Astronomy, K\"onigstuhl 17, D-69117 Heidelberg, Germany}
\altaffiltext{25}{Universit\'e Lyon 1, Observatoire de Lyon, Centre de Recherche Astrophysique de Lyon \\ \hspace*{0.5cm} and Ecole Normale Sup\'erieure de Lyon, 9 avenue Charles Andr\'e, F-69230 Saint-Genis Laval, France}
\altaffiltext{26}{Department of Physics and Astronomy, Texas Christian University, TCU Box 298840, Fort Worth, TX 76129, USA}
\altaffiltext{27}{Department of Physics and Astronomy, 203 Van Allen Hall, The University of Iowa, Iowa City, IA 52242-1479, USA}
\altaffiltext{28}{Department of Astrophysical Sciences, Princeton University, Princeton, NJ 08544, USA}
\altaffiltext{29}{Department of Astronomy, Case Western Reserve University, Cleveland, OH 44106, USA}
\altaffiltext{30}{Max–Planck–Institut f\"{u}r Astrophysik, Karl–Schwarzschild–Str. 1, D-85748 Garching, Germany}
\altaffiltext{31}{Departamento de Astrof\'\i sica, Universidad de La Laguna, E-38206 La Laguna, Spain}
\altaffiltext{32}{Shanghai Astronomical Observatory, Nandan Road 80, Shanghai 200030, China}
\altaffiltext{33}{Institute of Astronomy and Astrophysics, Academia Sinica, Taipei 106, Taiwan}
\altaffiltext{34}{National Astronomical Observatories of China, 20A Datun Road, Chaoyang District, Beijing 100012, China}
\altaffiltext{35}{Jodrell Bank Centre for Astrophysics, University of Manchester, Manchester M13 9PL, UK}
\altaffiltext{36}{Department of Physics and Astronomy, Macquarie University, NSW 2109, Australia}
\altaffiltext{37}{Australian Gemini Office, Australian Astronomical Observatory, PO Box 915, Sydney NSW 1670, Australia}
\altaffiltext{38}{Department of Physics, Yale University, 260 Whitney Ave, New Haven, CT 06520, USA}
\altaffiltext{39}{Physics Division, Lawrence Berkeley National Laboratory, Berkeley, CA 94720-8160, USA}
\altaffiltext{40}{Leibniz-Institut fuer Astrophysik Potsdam (AIP), An der Sternwarte 16, D-14482 Potsdam, Germany}
\altaffiltext{41}{Department of Physics \& Astronomy, University of Victoria, Victoria BC V8P5C2 Canada}
\altaffiltext{42}{Kapteyn Astronomical Institute, University of Groningen, Landleven 12, 9747 AD Groningen, The Netherlands}
\altaffiltext{43}{Department of Astronomy and Astrophysics, University of Toronto, 50 St. George Street, Toronto, Ontario M5S 3H4, Canada}